\documentclass[aps,pre,twocolumn,groupedaddress,showpacs]{revtex4-1}

\usepackage[utf8]{inputenc}
\usepackage[english]{babel}
\usepackage{amsmath,graphicx,enumerate,empheq}
\usepackage{epstopdf}

\usepackage{multirow}
\usepackage{hyperref}
\usepackage{color}

\usepackage{bm}
\usepackage[a4paper,includeheadfoot,margin=2.54cm]{geometry}
\usepackage{mathtools}

\usepackage{comment}

\newcommand{\U}{\mathcal{H}}
 
\begin{document}

\title{Optimisation of an active heat engine}

\author{Giulia Gronchi$^{1}$}
\author{Andrea Puglisi$^{2}$}
\affiliation{$^1$Dipartimento di Fisica, Universit\`a di Roma  Sapienza, P.le Aldo Moro 2, 00185, Rome, Italy \\$^2$Istituto dei Sistemi 
Complessi - CNR and Dipartimento di Fisica, Universit\`a di Roma Sapienza, P.le Aldo Moro 2, 00185, Rome, Italy }

\date{\today}

\begin{abstract}
Optimisation of heat engines at the micro-scale has applications in biological and artificial nano-technology, and stimulates theoretical research in non-equilibrium statistical physics. Here we consider non-interacting overdamped particles confined by an external harmonic potential, in contact either with a thermal reservoir or with a stochastic self-propulsion force (active Ornstein-Uhlenbeck model). A cyclical machine is produced by periodic variation of the parameters of the potential and of the noise. An exact mapping between the passive and the active model allows us to define the effective temperature $T_{eff}(t)$ which is meaningful for the thermodynamic performance of the engine. We show that $T_{eff}(t)$ is different from all other known active temperatures, typically used in static situations. The mapping allows us to optimise the active engine, whatever are the values of the persistence time or self-propulsion velocity. In particular - through linear irreversible thermodynamics (small amplitude of the cycle) - we give explicit formula for the optimal cycle period and phase delay (between the two modulated parameters, stiffness and temperature) achieving maximum power with Curzon-Ahlborn efficiency. In the quasi-static limit, the formula for $T_{eff}(t)$ simplifies and coincides with a recently proposed temperature for stochastic thermodynamics, bearing a compact expression for the maximum efficiency. A point - overlooked in recent literature - is made about the difficulty in defining efficiency without a consistent definition of effective temperature.
\end{abstract}


\maketitle

\section{Introduction}

In his talk “There is plenty of room at the bottom,” Feynman
envisioned a microscopic motor working at small scales, even at the
single-atom level~\cite{feynman59}. Such a motor would be the first
step required to achieve the ability of ``manipulating and controlling
things on a small scale''. Sixty years later, Feynman's idea has been
realised in several experiments and its theoretical implications have
been deeply analysed~\cite{blickle12,rossnagel14,martinez15}.

A motor is a device that delivers mechanical work, for instance by
pushing a weight in a given direction. Work is obtained by converting
a fraction of energy taken from  reservoirs: such a fraction
represents the motor's efficiency. In the microscopic world, the list
of available energy reservoirs is not substantially different from the
macroscopic scale, i.e.  mainly chemical or electro-chemical
reservoirs or chemically-induced heat reservoirs~\footnote{the
  macroscopic world has additional sources of energy, unfortunately -
  for our environment - of minor importance for the moment, such as
  those related to natural macroscopic flows, e.g. air and
  water}. In physics of course the choice of heat reservoirs is the
one that better stimulates theoretical research as it involves
translating principles of thermodynamics to small scales, far from the
thermodynamic limit~\cite{martinez17}. The challenge, with microscopic
heat engines, is to achieve optimal control of thermal
fluctuations, which are involved not only as energy source and sink
(as in macroscopic heat engines) but also spoil the stability and
reliability of the delivered work. Microscopic work, and therefore
efficiency, are in fact highly fluctuating
quantities~\cite{verley14}. The effect, also beneficial, of
fluctuations on motor efficiency is one of the most intriguing recent
discoveries in the field of stochastic
thermodynamics~\cite{pietzonka18}.

\subsection{Passive microscopic heat engine}

Microscopic heat engines have been at the center of theoretical and
experimental research in the last decade. They have been realised with
colloidal particles in optical traps, for instance in Stirling cycles
with isochoric and isothermal transformations~\cite{blickle12} and
Carnot cycles with adiabatic and isothermal
transformations~\cite{martinez15}. The realisation of adiabatic
passages (which would require complete isolation of the Brownian
particle) is obtained through a protocol where both characteristic
volume and temperature of the system are changed in such a way that
the entropy of the system is conserved~\cite{martinez15b}. Heat
engines have also been realised at the atomic scales, by manipulating
a trapped ion~\cite{rossnagel16}. The exploitation, at single-atom
level, of so-called quantum squeezed states has also been proposed as
a way to circumvent Carnot limit in efficiency~\cite{rossnagel14}. The
theoretical research on small-scale engines has involved also the
possibility of designing specific cycles with optimal power or
efficiency~\cite{schmiedl08}. An important challenge, in this field,
is going beyond the single-particle limit and achieve control/design
of systems made of a small number (e.g. $100-1000$) of particles,
which can be meaningful for biophysical applications~\cite{cerino16}.

A severe limitation against a straightforward experimental realisation
of a heat engine stems from the difficulty of controlling the
temperature with due precision, plagued by the unwanted development
of gradients and the presence of large relaxation
times~\cite{martinez17}. A typical workaround is to replace the high-temperature reservoir with a source of noise, e.g. an applied noisy
voltage as in~\cite{blickle12,martinez15}. A different fascinating
possibility is to consider engines made of a different kind of working
substances which stay at an {\em effective temperature} different
(typically higher) than the environment/solvent~\cite{puglisi2017temperature}. This can be achieved
by means of active particles, i.e. particles which are self-propelled,
for instance bacteria or sperms or active colloids (e.g. Janus
particles)~\cite{dileo16}. 

\subsection{Active heat engines}

Every active particle has its own internal
motor which induces, in the presence of a viscous solvent, a typical
speed $v_0$. Of course $v_0$ is unrelated to the thermal speed, that
is $v_0 \neq \sqrt{k_B T/m}$ where $T$ is the temperature of the
environment, $k_B$ is the Boltzmann constant and $m$ is the mass of
the particle. Most importantly - considering that active
micro-swimmers move through overdamped kinematics - their unconfined
diffusivity $D_a$ is typically much larger than molecular diffusivity $D$:
$D_a=\tau_a v_0^2 \gg D = k_B T/\gamma$, being $\tau_a$ the active
persistency time and $\gamma$ the viscous drag of the
particle~\cite{dileo16}. This consideration leads naturally to define
a diffusivity-based active temperature $T_D=\gamma D_a/k_B \gg
T$. The equilibrium limit $\tau \to 0$ is typically taken in such a way that $T_D \to T$, which requires $v_0^2 \to D/\tau_a$.

First studies and experiments demonstrating the possibility of
converting random self-propulsion into directed motion or work have
been realised in the realm of active
ratchets~\cite{dileonardo10,sokolov10,vizsnyiczai17,reichhardt17,pietzonka2019},
which are autonomous engines. In the most recent years several
proposals of cyclical heat engines have been done and in few cases
also experimentally realised.

In~\cite{krishnamurthy2016} a first example of Stirling engine (two
isotherms and two isochores) was obtained, where bacteria were
involved as bath and the central system was made of a trapped
colloidal particle: an external control of the solvent temperature was
reflected in a variation of the average speed (activity) of the
bacteria, measured through tracking the position fluctuations of the
colloidal particles. The authors verified that iso-thermal
transformations (compressions and expansions) were also
``iso-active'', i.e. activity did not depend appreciably upon the trap
stiffness. The advantage of such a bacterial bath was to achieve a
much larger range of effective (active) temperatures than in the
passive case.

The concept of an active Stirling engine was investigated in a more
recent theoretical study~\cite{zakine2017}. The authors show that the
performances of the engine depend upon what is the ``temperature''
which is kept fixed during the iso-thermal transformations: the two
considered candidates (here we set $k_B=1$) are $T_{var}=k \langle x^2 \rangle$, as
chosen in~\cite{krishnamurthy2016}, related to energy in a harmonic
potential of stiffness $k$, and the diffusion temperature $T_D$
defined above, as proposed in~\cite{wu2000}. At equilibrium (ie. for thermal particles at temperature $T$) of course
$T_{var}=T_D=T$. The authors consider several different models for the
bacterial bath (including non-Gaussian effects and/or temporal
correlations, i.e. persistency), concluding that if $T_{var}$ is kept
constant in iso-thermal transformations then the equilibrium limit for
efficiency (given by Carnot value) cannot be surpassed, while
different things may happen if $T_D$ is adopted for the iso-thermal
branches of the cycle.

Other authors~\cite{martin2018} have considered a different
theoretical model where the central particle is a self-propelled
particle (pushed by a random force with exponentially decaying
autocorrelation - with typical time $\tau_a$ - as in AOUP model, see
below) immersed in a bath of passive particles. The authors consider
both Stirling type engines (cyclical modulation of temperature and
stiffness at fixed $\tau_a$) and engines with modulation of $\tau_a$ and
stiffness, at fixed temperature. Of course the second case has not a
passive counterpart and therefore there is no direct way to compare
performances. It is important to stress that the authors, here, have
decided to connect the thermal bath temperature to the self-propulsion
speed, similar to what happens in~\cite{krishnamurthy2016}, making
more difficult to disentangle their contributions.

In~\cite{saha2019}
a new heat engine is proposed where a passive particle is trapped in a
harmonic potential with time-dependent stiffness and is put in contact
with a thermal bath in the first half of the cycle and with an active
bath (time-persistent noise) in the second half. In this paper the
relevance of $T_{var}$ as a sort of effective temperature
and a general equation for its evolution - for a broad class of choices
of the driving noise - is shown (an equation discussed in greater detail in the
later~\cite{holubec2020}).

In~\cite{kumari2020} a Stirling engine is considered where the central
substance is a particle that changes its nature during the cycle
itself, i.e. it is passive for three of the four steps, and is an
active Ornstein-Uhlenbeck particle (AOUP) during the fourth step which
is the isothermal compression. In this paper a higher efficiency (with
respect to the passive case) is claimed when activity is present, a
fact which evidently depends upon the chosen definition of efficiency.

In~\cite{ekeh2020} the authors consider a model with many active
Brownian particles (ABP, i.e. such that their self-propulsion velocity has
fixed magnitude and diffusing orientation). The external potential
(which also act on propulsion's orientation) has many parameters that
can be varied. The presence of many potential's parameters allows one
to design cycles without changing other properties (such as bath
temperature or properties of the activity). Efficiency appears to be
proportional to the extracted power and both are optimal together.

We conclude this overview of the very recent literature
with~\cite{holubec2020}, where the authors propose a general mapping
from an active heat engine to a passive heat engine. The mapping can
be made explicit when the confining potential is harmonic, and this
can be done for whatever model of self-propulsion is proposed: only
the auto-correlation of the self-propulsion affects the evolution of
the effective temperature. The authors give explicit examples of their
formalism using an ABP particle in a harmonic trap, where a
Stirling-like cycle is operated by tuning stiffness, temperature and
parameters of activity (both speed and persistency time). The
important point raised by the authors, also relevant for the
interpretation of the experiments in~\cite{krishnamurthy2016}, is that
the effective temperature may change also during the (apparently)
``isothermal'' transformations, because it depends upon all the
system's parameters.

\subsection{This paper}

Here, we propose a study which is  complementary to that done
in~\cite{holubec2020}. We consider AOUP particles which have a natural mapping to passive systems with an
active temperature, $T_a$(t), which in fact depends upon the potential's
parameters, in the low persistence limit. A recent study of entropy production for AOUP particles also revealed the existence of a Clausius relation that connects entropy changes with a an active heat flow divided by a temperature $T_{cl}(t)$~\cite{umb17,clau17}.  Our main point is that the proper effective temperature, $T_{eff}(t)$, relevant for the
thermodynamic behaviour of the engine, can be quite different from $T_a(t)$ and $T_{cl}(t)$,
as well as from other temperatures such as $T_{var}$, $T_D$ or $T_a$. Later, in Table~\ref{t:temp} we summarise some of the most used definitions of temperatures in this context.

The
knowledge of the correct effective temperature is crucial to optimize
the engine's performance, for instance of its delivered power. As an explicit application of this concept, we show how to achieve
maximum power by tuning $T_{eff}$ through the control of
self-propulsion speed, a possibility which has been experimentally
realised, recently, with light-controlled bacteria~\cite{vizsnyiczai17,arlt2019dynamics} and colloids~\cite{schmidt2019light}.

Let us summarise the structure of the paper. In Section II we
introduce a few standard thermodynamic tools which are useful for
finite-size/time thermodynamics (i.e. when models are stochastic and
the period of a cyclic transformation is not infinite): in this
Section we briefly discuss the correct definition of heat flow coming
from the high-temperature thermostat, which is important already for
non-active systems, and becomes even more important for active heat
engines where the temperature is not directly under control.  In Section
III we present the passive and active models investigated in the
paper, showing the mapping that brings them to be equivalent from the
point of view of heat engine performances~\cite{holubec2020}. In
Section IV we study the periodic heat engine model with only passive
particles, particularly in the limit of linear irreversible
thermodynamics, deriving a few results which are
complementary to those given for the same model
in~\cite{brandner2015}. In Section V we show how to transfer the
knowledge of the passive engine to optimise the active one. In the
last section we draw conclusions and perspectives.

\section{Work and heat for microscopic (passive) heat engines}

In the context of microscopic engines, basic thermodynamic concepts,
such as heat and work, need a definition in terms of stochastic
quantities, even if only averages are needed. For simplicity we
consider models with a single (passive or active) particle in a
solvent fluid which is viscous enough to make inertia negligible. The
particle therefore obeys dynamical equations which result from some
external potential $\U({\bf x},{\bf \lambda}_t)$, thermal fluctuations
and self-propulsion (in the active case). An external agent can
control the parameters ${\bf \lambda}_t$ of the potential and also the
properties of the thermal bath and the active self-propulsion: in
doing so, it performs or extracts work. In stochastic thermodynamics the
definition of stochastic work injection rate (or injected power), is
related to the variation of the external potential, i.e.~\cite{seifertrev}:
\begin{equation}
	\dot{W}= \sum_i\frac{ \partial \U}{\partial \lambda_i} \lambda_i= \frac{\partial \U} {\partial t}.
	\label{eq:lavoro}
\end{equation}
A way to verify the consistency of this relation with
thermodynamics~\cite{jarzynski2007comparison} is to study what happens
in the quasi-static limit when no active forces are present. This
amounts to consider the limit $t_F \to \infty$ (where $t_F$ is the
duration of the transformation). In this limit the system is
constantly close to its equilibrium distribution $p_0$ at given
potential's parameters ${\bf \lambda}_t$, where
\begin{equation}
	p_0 = \frac{1}{\mathcal{Z}({\bf \lambda}_t)} e^{-\beta \U({\bf x},{\bf \lambda}_t)},
\end{equation}
with
\begin{equation}
\mathcal{Z}({\bf \lambda}_t)=\int d{\bf x} e^{-\beta \U({\bf x},{\bf \lambda}_t)} 
\end{equation}
and $\beta$  the inverse temperature.
In particular one has
\begin{equation}
  \partial_t p_0 = -\beta \big[\partial_t \U \big] p_0 - \frac{\partial_t \mathcal{Z}(\lambda_t)}{\mathcal{Z}(\lambda_t)} p_0 = \mathcal{O}
\Big(\frac{1}{t_F}\Big),
\end{equation}  
and therefore, according to definition~\eqref{eq:lavoro}, the average work done in the quasi-static limit
corresponds - as expected - to the free energy variation
\begin{equation}
  \int  \big[ \partial_t \U \big] p_0 d{\bf x} =  \partial_t \Big[- \frac{1}{\beta} \ln \mathcal{Z}(\lambda_t)\Big]+ \mathcal{O}\Big(\frac{1}{t_F}
\Big).
  \end{equation}

Instantaneous heat exchange is defined for complementarity from work, i.e. 
\begin{equation} \label{heat}
  \dot{Q}=\frac{dH}{dt}-\dot{W},
\end{equation}
so that the first principle is guaranteed. When the parameters
$\lambda_t$ are tuned according to a cyclical protocols,
i.e. $\lambda_{t+t_{cycle}}=\lambda_t$ with $t_{cycle}$ the machine
period, most of the models and the initial conditions lead to a limit
cycle where averages are periodic with the same period $t_{cycle}$. It
is then meaningful to consider the average work integrated in a period
\begin{equation} \label{eq:wp}
W_p = \int_{t_0}^{t_0+t_{cycle}}dt \langle \dot{W} \rangle,
\end{equation}
(where $t_0$ is a time large enough to consider the system in the
limit cycle). We recall that $W_p$ - in our notation - must be negative to have a working machine. 

\subsection{Adsorbed heat}

For the purpose of computing the engine's efficiency, it is crucial to define a
measure of energy consumption through heat:
\begin{equation} \label{eq:qh}
Q_h = \int_{t_0}^{t_0+t_{cycle}} dt w_{ads}(t) \langle \dot{Q} \rangle,
\end{equation}
where $w_{ads}(t) \in [0,1]$ is a (periodic) weighting function that
discriminates how much heat is to be considered as energy gain, while the remaining fraction $1-w_{ads}(t)$ is to be
considered as dissipation.

We believe that a good choice (and a good understanding) of
$w_{ads}(t)$ - even in the framework of passive, not active, particles
- deserves a brief discussion, since it seems that there is not
unanimous agreement about it in the  literature, even the recent one. In Carnot's
original heat engine, there are two well defined thermostats, i.e. the
one at high temperature $T_h$ and the one at low temperature $T_c$,
with adiabatic connections: then, heat is only adsorbed when in
contact with $T_h$ and is only released when at $T_c$. This means
that one can safely set 
\begin{equation} \label{theta}
w_{ads} (t) = \Theta[\langle \dot Q \rangle(t)],
\end{equation}
where $\Theta$ is the Heaviside theta
function~\cite{martinez15,martinez15b}.  However, in more general
cases, definition~\eqref{theta} has drawbacks. For instance the
Stirling engine - even in the quasi-static limit - exchanges heat
during the isochoric branches of the cycle, when in contact with
intermediate values $T_c < T(t) < T_h$: if Eq.~\eqref{theta} is adopted, then $Q_h$ takes a contribution
also from one of the isochores and the quasi-static
efficiency is, in general, smaller than the Carnot
one~\cite{blickle12,zakine2017}. Such a definition is also problematic
from the conceptual point of view, since $Q_h$ should be related to
 heat flowing from a thermostat to a different one, however Eq.~\eqref{theta} gives $Q_h>0$ also when there is a single
thermostat, i.e. when $Q_h$ comes from the same thermostat which - considering the whole transformation - dissipates heat: this happens in several examples with a
time-dependent Hamiltonian (periodically modulated potential energy)
at a constant temperature. Notwithstanding this drawbacks, 
definition~\eqref{theta} is frequently
adopted~\cite{blickle12,krishnamurthy2016,zakine2017,martin2018}~\footnote{A recent different definition for $Q_h$ has been proposed in the context of active engines coupled both with a steady active bath and a steady thermal bath which are of course at different temperatures~\cite{fodor2021active}: in that case a cyclical engine can be obtained by tuning in time {\em two} parameters of the external potential and the proposed definition of adsorbed heat is the whole heat exchanged with the active bath, which is positive on average. In our opinion this choice should be debated, as it implies that no heat is dissipated in the active bath itself, a fact which we are challenging. }.

An alternative recipe is offered
in~\cite{brandner2015,cerino16}, adopted by us in the present study:
\begin{equation}
w_{ads}(t)=\frac{\beta(t)-\beta_c}{\beta_h-\beta_c}, \label{seif}
\end{equation}
where $\beta(t)=1/T(t)$, $\beta_{c(h)}=1/T_{c(h)}$. This definition weighs more heat (whatever its sign) coming from higher temperature. Eq.~\eqref{seif}
is justified by splitting the entropy production into two
contributions that depend upon two different thermodynamic forces, one
related to the variation of potential energy (which generates work)
and one related to the variation of temperature (which generates a
heat flux $Q_h$ going through the system from high to low
temperature)~\cite{brandner2015}. Such a recipe also
guarantees that in the quasi-static limit the Carnot efficiency
$\eta_c=1-T_c/T_h$ is always reached, including the Stirling engine~\footnote{With this definition the entropy produced in a period due to work and heat flux are equal to $S_{prod,W}=\beta_c W_p$ and $S_{prod,h}=Q_h(\beta_c-\beta_h)$ respectively~\cite{brandner2015}, so that $\eta=-W_p/Q_h = S_{prod,W}(\beta_h-\beta_c)/(S_{prod,h}\beta_c)$. In the quasi-static limit $S_{prod,W}/S_{prod,h}=-1$,  which leads to the Carnot efficiency $\eta=1-\beta_h/\beta_c=\eta_c$.}. 

Once work and adsorbed heat are defined, one can define the average power
\begin{equation}
P=\frac{W_p}{t_{cycle}}, \label{eq:P}
\end{equation}
and the average efficiency
\begin{equation}
\eta=-\frac{W_p}{Q_h}.  \label{eq:eta}
\end{equation}

\subsection{Active-passive equivalence}
  
A fundamental observation can be done for harmonic potentials, i.e. when
\begin{equation}
\U(t) = \frac{1}{2} k(t) x^2(t).
\end{equation}
For such a choice, work (and total heat which is the opposite of work, in a period)  are known through the knowledge of $\sigma(t)=\langle x^2
\rangle$ and $k(t)$ only~\cite{holubec2020}:
\begin{equation}
W_p= \int_{t_0}^{t_0+t_{cycle}} dt \frac{1}{2} \dot{k} (t)\sigma(t), 
\label{eq:w}
\end{equation}
The same holds true for the instantaneous total heat $\langle \dot Q \rangle$, see Eq.~\eqref{heat} as well as for its integral over a period.
On the contrary, and at variance with what concluded in~\cite{holubec2020}, adsorbed heat
\begin{equation} \label{eq:q}
Q_h = \int_{t_0}^{t_0+t_{cycle}} dt w_{ads}(t) \frac{1}{2} k(t) \dot\sigma(t)
\end{equation}
in general {\em does not} depend only upon $\sigma(t)$ and $k(t)$, since the definition of $w_{ads}(t)$ could depend upon other parameters, for instance - in the definition adopted by us, Eq.~\eqref{seif} - it depends upon $T(t)$.

\section{Particle models}

In this section we discuss the model of active (and passive) particle and the
adopted cycle for the heat engine we want to study. We stick to a
harmonic potential with time-dependent stiffness and we adopt the AOUP
model for self-propulsion. 

\subsection{Passive model}

As a reference, we consider first an overdamped passive particle with
time-dependent diffusivity $D(t)= \frac{k_B T(t)}{\gamma}$. We choose units such that
the viscous drag $\gamma$ and the Boltzmann constant $k_B$ are both set to $1$. The model
then reads
\begin{equation}
dx(t)= - k(t) x(t) dt + \sqrt{2 T(t)} dw(t),
\label{eq:OU1}
\end{equation}
where $x(t)$ is particle's position at time $t$, $k(t)$ is the time-dependent harmonic stiffness, and $dw(t)$ is the infinitesimal increment of the Wiener process. The model has been studied in details in~\cite{schmiedl08}.  The model
has a Gaussian propagator and - in the absence of drifts and initial
displacements - its dynamics is fully described by the variance of the
position $\sigma(t)$ which obeys:
\begin{equation}
\dot{\sigma}(t)= - 2 k(t) \sigma(t) + 2 T(t) .
\label{eq:sigma}
\end{equation}

In the quasi-static limit of $t_{cycle} \to \infty$, one can change 
the time-variable in Eq.~\eqref{eq:sigma} defining $s=\frac{t}{t_{cycle}}$ 
(so that in $s$ the period is $1$) and obtaining
\begin{equation} \label{qssigma}
- 2 k(s)\sigma(s) + 2 T(s) = \mathcal{O}(1/t_{cycle}).
\end{equation}
which of course in the limit $t_{cycle} \to \infty$ gives $\sigma(t) = \frac{T(t)}{k(t)}$.

\subsection{AOUP active model}

The AOUP model is considered a good description for a colloid in a
bath of swimmers such as bacteria~\cite{MBGL15}. This model has some
properties in common with the ABP model, for instance the
exponentially decaying time-correlation of the cartesian components of
the self-propulsion force (but ABP has non-Gaussian fluctuations and therefore 
it is more complicate to get analytical results). The AOUP model has the 
advantage of being more
accessible to calculations, in particular it has a very well studied
approximation in the passive limit (see below), where -- in the case
of constant parameters -- it is mapped into a passive model. We will see, however, 
how this mapping is not the proper one to understand the performance of this model as a heat engine.

We consider non-interacting AOUP particles (in one dimension) in the
hypothesis of large viscosity, that is inertia is neglected and the dynamics
is overdamped. We also neglect the thermal noise (which has a small effect
with respect to activity), writing :
\begin{equation} \label{eq:aoup}
\begin{cases}
dx(t) = [- k(t) x(t)  + f_a(t)] dt  \:\: \\
df_a(t) = -\frac{1}{\tau_a} f_a(t) + \frac{{2 D_a}^{1/2}}{\tau_a} dw(t). 
\end{cases}
\end{equation} 
The role of self-propulsion is played by
$f_a(t)$ which is a coloured noise with exponentially decaying
time-correlation:
\begin{equation}
\langle f_a(t) f_a(s) \rangle = \frac{D_a}{\tau_a} e^{-(t-s)/\tau_a},
\end{equation}
modelling a force which remains persistent for a time of order $\tau_a$.
The self-propulsion model has two parameters: the persistence time
$\tau_a$ and the active diffusivity $D_a$. In the absence of external
potential and if $D_a$ is constant, at large times $t \gg \tau_a$,  the particle displays normal
diffusivity with coefficient $D_a$. From it one can define the average
self-propulsion speed $v_0(t)=\sqrt{D_a(t)/(\tau_a)}$. The passive model is
recovered taking $\tau_a \to 0$ and $v_0(t) \to \infty$ with $D_a \to
D$. The more general case we consider is a time-dependent $D_a(t)$ or
$v_0(t)$, a situation which has been demonstrated experimentally, for
instance in~\cite{vizsnyiczai17}.

The model, being the external potential harmonic, is again a Markovian
diffusive model with Gaussian propagator. Again, in the absence of
drifts and initial displacements, the dynamics is fully described by
the entries of the covariance matrix, that obey the following system
of coupled equations:
\begin{subequations}
  \begin{align}
\frac{ d{ \langle f_a^2 \rangle }}{dt} &= - \frac{2}{\tau_a} \langle f_a^2 \rangle + \frac{2 v_0^2}{\tau_a} \label{ff} \\
\frac{d { \langle xf_a \rangle }}{dt} &= \langle f_a^2 \rangle - \frac{1+\tau_a k}{\tau_a} { \langle xf_a \rangle } \label{xf}\\
\frac{d { \langle x^2 \rangle }}{dt} = \dot{\sigma}=& - 2 k { \langle x^2 \rangle } + 2 { \langle x f_a \rangle }
\end{align}
\label{sis1}
\end{subequations}
The latter is equivalent to Eq. (15) of~\cite{holubec2020} and defines
- for direct comparison with Eq.~\eqref{eq:sigma} - the effective
temperature
\begin{equation}
T_{eff}(t) = \langle xf_a \rangle (t),
\end{equation}
that is the temperature of the passive system which gives the same $\sigma(t)$ and therefore the same delivered work or power. Remarkably, this expression of $T_{eff}$ is directly proportional to the kinetic temperature $T_{kin}=m \langle v^2 \rangle/d$ (where $d$ is the dimensionality) recently calculated in a static harmonic potential for an inertial AOUP model~\cite{caprini21}. Interestingly, it is also proportional to the so-called swim pressure~\cite{takatori2014swim,winkler2015virial}. 

In the present case, $T_{eff}(t)$ obeys quite a simple differential equation,  i.e. (from joining together equations~\eqref{ff} and~\eqref{xf})
\begin{multline} 
\label{mappingaoup}
\ddot{T}_{eff} +\left(\frac{3+\tau_a k(t)}{\tau_a} \right) \dot{T}_{eff}+\\
+\left[\dot{k}(t)+2\frac{1+ \tau_a k(t)}{\tau_a^2}\right]T_{eff}(t)-2  
\frac{v_0^2(t)}{\tau_a}=0,
\end{multline}
which represents a central result of this paper.
In the passive limit $\tau_a \to 0$ with $\tau_a v_0^2(t) \to D_a(t)$
the above equation gives the correct expectation $T_{eff}(t) \to T_D(t)$. However, already at first order in $\tau_a$, one has
$T_{eff}(t) \neq T_D$.

When the parameters are constant or vary very slowly ($\omega=2\pi/t_{cycle} \to 0$
as done in Eq.~\eqref{qssigma}), the system reaches a (steady or
quasi-static) state where
\begin{subequations}
\begin{align}
\langle f_a^2 \rangle(t) &= v_0^2 (t)\\
\langle xf_a \rangle (t) &=T_{eff}(t)=\frac{v_0^2 (t)\tau_a}{1+ \tau_a k(t)} \label{teffqs}\\
\langle x^2 \rangle (t)&=\sigma(t)=\frac{ v_0^2 (t)\tau_a}{k(t)[1+\tau_a k(t)]}=\frac{T_{eff}(t)}{k(t)}. \label{sigmastatic}
\end{align}
\end{subequations}
It is useful to stress that, even in the quasi-static limit (very slow
transformations), $T_{eff}(t) \neq T_D(t)$ if $\tau_a>0$,
i.e. if the system is active. In other terms, even with very slow
transformations, an active system has a different thermodynamics with respect to a
passive one. Interestingly in the recent literature a temperature equal to $T_{eff}$ in the quasi-static limit, $T_{cl}=\frac{v_0^2 \tau_a}{1+ \tau_a k}$ has  shown to bear thermodynamic properties, as it underlies a Clausius relation for the entropy change of active particles~\cite{umb17,clau17}.

We conclude this section remarking that the meaning of the effective
temperature $T_{eff}(t)$ is twofold: first it is the temperature which
- replaced into $T(t)$ in the passive model Eq.~\eqref{eq:OU1} - gives
the same evolution of $\sigma(t)$ in the presence of the same protocol
$k(t)$, which guarantees that work, heat and power are
exactly the same; second, in the steady state (and in quasi-static
transformations) it rules the Boltzmann distribution $p(x) \sim
e^{-k(t)x^2/T_{eff}(t)}$.

\subsection{An approximation for small persistency}

\label{sec:fox}
It is easy to show~\cite{MBGL15} that the model in~\eqref{eq:aoup} can
be mapped, without any approximation to a Klein-Kramers model with an
effective mass $\tau_a$, a harmonic force with an effective stiffness
$\dot k +\frac{k}{\tau_a}$ and a viscous bath with effective drag
coefficient $1+\tau_a k(t)$:
\begin{equation}
\begin{cases}
dx= v dt \\ dv = - \frac{\Gamma(t)}{\tau_a} v dt - \left(\dot k
+\frac{k}{\tau_a}\right) xdt + \sqrt{\frac{2 v_0^2(t)}{\tau_a}} dw
\end{cases},
\label{eq:aoup_UD_1d}
\end{equation}
with
\begin{equation}
\Gamma(t) = 1 + \tau_a \partial_{x}^2\U = 1+\tau_a k(t).
\end{equation}
Again, the model has a Gaussian
propagator and its dynamics is described by the coefficient of the covariance matrix:
\begin{equation}
\begin{cases}
\frac { d\langle v^2 \rangle }{dt} = - \frac{2 \Gamma(t)}{\tau_a} \langle v^2
\rangle + 2 \frac{\tau_a \dot k+ k}{\tau_a} \langle xv \rangle +
\frac{2v_0^2}{\tau_a} \\ 
\frac { \langle xv \rangle }{dt} = \langle v^2
\rangle + \frac{\Gamma(t)}{\tau_a} \langle xv \rangle- \frac{\tau_a \dot k+
  k}{\tau_a} \langle x^2 \rangle \\ 
  \frac { \langle x^2 \rangle }{dt} = 2 {
  \langle xv \rangle }
\end{cases}
\label{sis2}
\end{equation}
whose steady (or quasi-static) state reads
\begin{equation}
\begin{cases}
\langle v^2 \rangle (t)=\frac{v_0^2(t)}{\Gamma(t)} =\frac{T_{cl}(t)}{\tau}\\
\langle xv \rangle =0\\
\langle x^2 \rangle(t) =\frac{v_0^2(t) \tau_a}{k(t) \Gamma(t)}=\frac{T_{cl}(t)}{k(t)}.
\end{cases}
\end{equation}

In the limit of $\tau_a \to 0$, the model in
Eqs.~\eqref{eq:aoup_UD_1d} can be approximated by a heuristic
procedure, equivalent to {\em overdamping}, where ``inertia''
(i.e. $dv$ ) is neglected. This procedure generalises to the case of
time-dependent parameters the so-called UCNA expansion~\cite{hanggi87,hanggi95}) and (for the case of a harmonic potential) gives
\begin{equation} \label{eq:fox}
dx = - \frac{\tau_a \dot k + k}{1+\tau_a k} x dt + \sqrt\frac{2 v_0^2 \tau_a}{(1+\tau_a k)^2} dw.
\end{equation}
For simplicity, in the rest of the paper we call this model "dynamical UCNA"
approximation. We notice that  it is equivalent to passive model, Eq.~\eqref{eq:OU1}, and therefore has variance satisfying Eq.~\eqref{eq:sigma}, with $k(t)$ and $T(t)$ replaced by
\begin{subequations}
\label{mappingfox}
\begin{align} 
  k_a(t)&=\frac{\tau_a \dot k(t) + k(t)}{1+\tau_a k(t)},\\
  T_a(t)&=\frac{v_0^2(t)   \tau_a}{(1+\tau_a k(t))^2}.
\end{align}
\end{subequations}
We highlight that $T_a(t) \neq T_{eff}(t)$, even at first order in
$\tau_a$. Of course in the passive limit ($\tau_a \to 0$ and $v_0^2(t)
\tau_a \to D_a(t)$) both temperatures $T_a(t)$ and $T_{eff}(t)$ 
go to $T_D(t)$.


In the steady or quasi-static regime (constant or very slowly varying
$k(t)$ and $v_0(t)$), $T_a(t)$ and $T_{eff}(t)$ are still different, even at first order in $\tau_a$:
\begin{subequations}
\begin{align} \label{ineq}
T_{eff}(t) &\approx \tau_a v_0^2 \left[1-k(t)\tau_a+\mathcal{O}(\tau_a^2)\right]\\
T_a(t) &\approx \tau_a v_0^2 \left[1-2k(t)\tau_a+\mathcal{O}(\tau_a^2) \right]\\
\end{align}
\end{subequations}
However $\sigma(t)=T_a(t)/k_a(t)$ coincides with that in Eq.~\eqref{sigmastatic}. For
small $ \tau_a$, it takes the form
\begin{equation}
\sigma=\frac{v_0^2 \tau_a}{k}(1-\tau_a k)+\mathcal{O}(\tau_a^3).
\end{equation}

It is important to understand that the "passive" problem with
parameters $k_a(t)$ and $T_a(t)$ is not {\em thermodynamically}
equivalent to our original active problem, since the work (and therefore power) of the original problem must be evaluated
against the original stiffness $k(t)$ and not against
$k_a(t)$. Therefore the analogy appearing in Eq.~\eqref{eq:fox} cannot
be immediately used for optimisation purposes. 
In Table~\ref{t:temp} we resume the main definition of temperatures used in this paper. We also summarise, in Table~\ref{t:lim}, the important physical limits which can be considered when discussing active heat engines.

\begin{table}
\begin{tabular}{|c|c|c|}
Name & Definition & Application \\ \hline
$T_D$ & $\gamma D_a/k_B$  & free diffusion~\cite{dileo16} \\ 
$T_{var}$ & $k\langle x^2 \rangle/k_B$  &  steady states~\cite{krishnamurthy2016} \\
$T_{kin}$ & $ m \langle v^2 \rangle/ k_B$ & steady states with inertia~\cite{caprini21} \\
$T_a(t)$ & $\frac{\gamma}{k_B} \frac{v_0^2(t)   \tau_a}{[1+\tau_a k(t)]^2}$ & dynamical UCNA~\cite{MBGL15} \\
$T_{cl}(t)$ & $\frac{\gamma}{k_B} \frac{v_0^2(t)   \tau_a}{1+\tau_a k(t)}$ & \begin{tabular}{@{}c@{}} Clausius relation~\cite{umb17,clau17}\\ (and $\lim_{\omega \to 0} T_{eff}(t)\;$)\end{tabular}  \\
$T_{eff}(t)$  & see Eq.~\eqref{mappingaoup} & heat engines~\cite{holubec2020}.
\end{tabular}
\caption{\label{t:temp} A table with the main definitions of effective temperatures used in the context of active particle models (in one dimension). We recall that  $\gamma$ is the viscous drag coefficient and $k_B$ the Boltzmann constant, both set to $1$ throughout the paper.}
\end{table}

\begin{table}
\begin{tabular}{|c|c|}
Name & Definition  \\ \hline
Passive limit & $\tau_a \to 0$, $v_0 \to \infty$, $\tau_a v_0^2 \to D_a$  \\ 
Quasi-static limit & $t_{cycle} \to \infty$ ($\omega \to 0$)\\
Linear limit & cycle amplitude ($\epsilon$, $\epsilon_k$, $\epsilon_T$) $\to 0$.
\end{tabular}
\caption{\label{t:lim} A table with the three important physical limits which can be considered when discussing an active heat engine (the definition of cycle amplitudes $\epsilon_k$, $\epsilon_T$ or $\epsilon$ are given in Section IV.A).}
\end{table}

\subsection{Possible strategies for optimizing the active heat engine}

We have shown that the active engine with parameters
$\tau_a,k(t),v_0(t)$ is equivalent, for the purpose of both the
evolution of $ \sigma(t)$ and the computation of work, to a passive
engine model defined in Eq.~\eqref{eq:OU1} with parameters $k(t)$ and
$T_{eff}(t)$ obeying Eq.~\eqref{mappingaoup}. Such an equivalence
allows to transfer results coming from the study of the passive model
to the active heat engine.

Note that the passive model has maximum efficiency (using for
$w_{ads}(t)$ the definition of~\cite{brandner2015}, that is Eq.~\eqref{seif}) given by the
Carnot efficiency in the quasi-static limit:
\begin{equation}
\eta \le \eta_c = 1-\frac{T_{min}}{T_{max}} \label{maxeff}
\end{equation}
where $T_{min}$ and $T_{max}$ are the minimum and maximum of $T(t)$
respectively.

In the active case this limit holds for the ``equivalent'' efficiency,
but one must take $T_{min}$ and $T_{max}$ as the minimum and maximum
of $T_{eff}(t)$ given by Eq.~\eqref{mappingaoup}.  The efficiency of
an active heat engine, however, is not an univocal concept. It
depends, through Eq.~\eqref{eq:eta}, upon the definition of $Q_h$,
which is already ambivalent for passive particles (see discussion at
the end of Section II) and is even more ambivalent for active ones,
since it could rely on the adopted choice of effective temperature.
Our point of view is that the choice of $T_{eff}$ for reference
temperature, together with the choice illustrated
in~\cite{brandner2015} for $Q_h$ (that is using Eq.~\eqref{seif},
detailed in Section IV.B), guarantee that $\eta$ reaches the Carnot
efficiency in the quasi-static limit, for any choice of the other parameters (including activity). Therefore, it is a meaningful
figure of merit, in the sense that it makes clear how far is the
machine from the maximum deliverable power. Of course a more severe
measure of efficiency could be considered, where $Q_h$ includes the
energy spent to feed the active particles, but this is of course out
of the scope of the paper (see discussion in~\cite{holubec2020}). 

\section{Discussion of the passive heat engine}

The optimisation of the passive model has been first studied in~\cite{schmiedl08} where a specific
Carnot-like protocol (two isothermal and two adiabatic) is considered and 
optimisation is done towards the maximum power  at fixed minimum/maximum
$\sigma$. A study of the same model within the framework of linear irreversible thermodynamics~\cite{izumida,callen} is presented
in~\cite{brandner2015}: in that study the Onsager coefficient
relative to the passive model for cyclical protocols $k(t)$
and $T(t)$ - undergoing small variations - are given, with formula for
efficiency and power as function of the parameters of the
model. Optimisation is done by fixing efficiency and the temperature protocol and looking for
the optimal stiffness protocol producing maximum power. In this Section we  consider a class
of harmonic protocols with phase shift (between stiffness and
temperature), investigating the more common question of the efficiency
at maximum power which - in this protocol class - results to be
equivalent to the Curzon-Ahlborn formula~\cite{curzon1975}.

In this paper we consider $k(t)$ and $T(t)$ to be periodic functions
with period $t_{cycle}$ corresponding to an angular frequency $\omega
= 2\pi/t_{cycle}$. The maximum variation of $k(t)$ and $T(t)$ are
proportional to $\epsilon_k$ and $\epsilon_T$ respectively.  In some situations we  consider $\epsilon_k
\propto \epsilon_T \propto \epsilon$.

Equation ~\eqref{eq:sigma} has a formal solution
\begin{equation} \label{formalsigma}
	\sigma(t) = \biggl[ \int_0^t e^{K(t')}  \: 2T(t') dt' + \sigma(0) \biggr] e^{-K(t)}
\end{equation}
where $\dot K(t) = 2k(t)$.  In general, whatever the initial variance
$\sigma(0)$, given the periodic protocols described above one observes
a relaxation of Eq.~\eqref{formalsigma} towards a limit cycle.  We
assume that this relaxation is achieved within a time $t_0$ (typically
a few periods are sufficient). Power and efficiency of the model, are
computed through integration of Equations~\eqref{eq:w}
and~\eqref{eq:q} with $\sigma(t)$ given by the solution
in~\eqref{formalsigma}. This task can be non trivial to be processed
analytically, even for simple protocols such as sinusoidal
functions. We resort to numerical integration of differential equations for $\sigma(t)$, $W_p$ and $Q_h$~\footnote{Our numerical scheme is a classical 4-th order Runge-Kutta integrator with time-step $dt=10^{-3}$ for the passive system and $dt=10^{-4}$ for the active one. } and, to get analytical formula, to the linear response regime,
i.e. when the amplitude of variations of $k(t)$ and
$T(t)$ is small.

\subsection{Qualitative picture of the cycle thermodynamics}
\label{sec:qualitative}

To get a first qualitative picture it is useful to set the protocols to simple sinusoidal functions (with same relative amplitude):
\begin{align} 	\label{protocols_easy}
k(t) &= k_0 + \epsilon_k \sin(\omega t) \\
T(t) &= T_c + \epsilon_T \frac{1-\cos(\omega t)}{2}.
\end{align}

\begin{figure}[h!]
	\centering
	\includegraphics[width=\columnwidth]{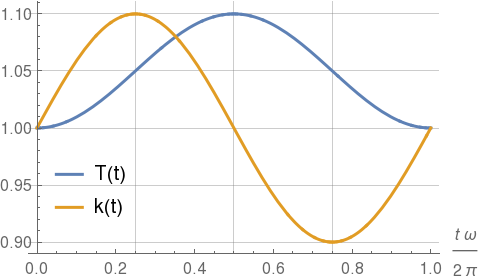}
		\caption{A sketch of the stiffness and temperature protocols used in Section IV.A. In Section IV.B we consider a phase shift (denoted with $\phi$) with respect to the case shown in the figure, for $T(t)$, which can be adjusted to optimise the delivered power. In this plot we have chosen $T_c=k_0=1$ and $\epsilon_k=\epsilon_T=0.1$\label{fig:prot}
}
\end{figure}

The protocol is illustrated in Fig.~\ref{fig:prot}. The stiffness and temperature variations are out of phase by a fourth of a period: the temperature maximum is synchronised with the instant where the expansion (decreasing stiffness) is fastest. This choice is inspired by the classical idealised Stirling engine. In the discussion of the linear response regime, below, we show that this choice is not optimal (i.e. a slightly different lag between temperature and stiffness can be found to increase delivered power), a fact  rarely discussed.

\begin{figure}[h!]
	\centering
	\includegraphics[width=\columnwidth]{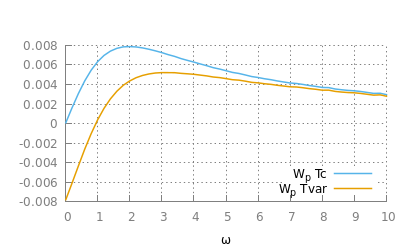} \\
	\includegraphics[width=\columnwidth]{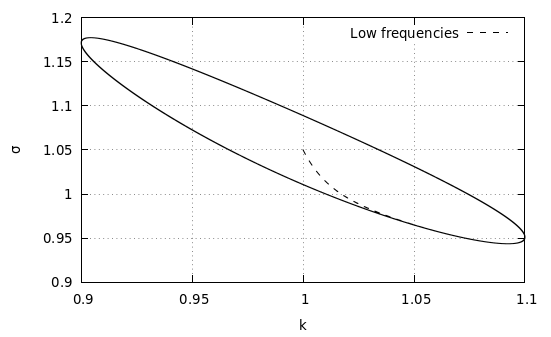}  	
	\includegraphics[width=\columnwidth]{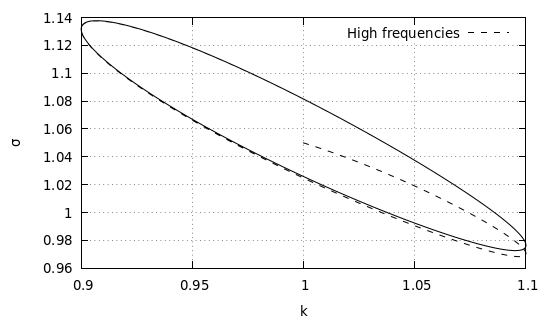}		
	\caption{Study of the passive engine.  A:  work  per cycle $W_p$ in the case at $\epsilon_T=0.1$ (blue curve, constant $T(t)=T_c$) and in the case $\epsilon_T=0.1$ (yellow curve, variable temperature). B and C: the Clapeyron plane and the different senses of rotation for two different frequencies of the engine cycle:  $\omega=0.2$ (B) and $\omega=2$ (C). Parameters: in all plots $k_0=1$, $\epsilon_k=0.1$ and $T_c=1$. \label{fig:1}
}
\end{figure}

In Figure~\ref{fig:1} we summarise the  behaviour of the passive heat engine with protocol given by Eqs.~\eqref{protocols_easy}. Frame A shows the total work per cycle $W_p$  as a function of $\omega$ for $\epsilon_k=0.1$ (with $k_0=1$ and $T_c=1$). The blue curve is for $\epsilon_T=0$ i.e. when the temperature {\em does not} change during the cycle (it is constant at $T=T_c$): the work in a period is never negative, a fact which is consistent with expectation from thermodynamics, i.e. there is no way to extract work from a single thermostat.  Moreover, in the quasi-static limit $\omega \to 0$ one has $\sigma(t)=T/k(t)$ at each time. In this limit one gets $W_p=0$ because the curve in the $k,\sigma$ plane goes back and forth along the same route and the enclosed area is empty.  As soon as a temperature variation is introduced, as seen in the yellow curve computed for $\epsilon_T=0.1$, the work may become negative i.e. there can be a positive power output so that the system behaves as a heat engine. This occurs at small frequencies (including the limit $\omega \to 0$), while at high frequency the work comes back to be positive and the machine stops acting as an engine. In frames B and C we show what happens in the plane $k,\sigma$. It is seen directly from Eq.~\eqref{eq:w} that a negative (= produced) work occurs when the limit cycle in that plane is swept in anti-clockwise direction, as it is observed for low frequencies (frame B) and opposite to high frequencies (frame C). 

The facts observed above can be understood analytically in the small perturbation limit, computing an approximated expression for $\sigma(t)$, even before going to the full linear response treatment discussed in the next section. For this purpose we consider two simplified situations: 
1) a situation where only the stiffness is perturbed so that  $\epsilon_k = k_0 \epsilon$ and $\epsilon_T=0$;
2) a situation where we assume that the two perturbations (stiffness and temperature) are similar, more precisely we set $\epsilon_k = k_0 \epsilon$ and $\epsilon_T=\epsilon T_c$.
In both cases  we set $\sigma(t)=\sigma_0(t) + \epsilon \sigma_1(t)$
and then replace it in the expanded Eq.~\eqref{eq:sigma}, equating equal powers in $\epsilon$, concluding by dropping terms with powers of $\epsilon$ larger than $1$.

In the first situation ($\epsilon_k = k_0 \epsilon$ and $\epsilon_T=0$) we get
\begin{multline}
	\frac{\sigma}{\sigma_s} = 1 + \epsilon \frac{2k_0}{\omega^2+4 k_0^2} \left[  -2 k_0 \sin(\omega t)   + \omega\cos(\omega t) \right]
	\label{eq:sigmaeps0}
\end{multline}
where we have defined the static variance $\sigma_s=T_c/k_0$ (the formula given here are valid if $0<\sigma_s<\infty$). Power adsorbed in this case reads
\begin{equation}\label{workeasy0}
P=\epsilon^2 k_0 T_c \frac{\omega^2}{2(4 k_0^2+\omega^2)},
\end{equation}
which is always non-negative, meaning that with $\epsilon_T=0$ this machine cannot do useful work, but only adsorb it.

In the second situation ($\epsilon_k = k_0 \epsilon$ and $\epsilon_T=\epsilon T_c$), instead, we get:
\begin{multline}
	\frac{\sigma}{\sigma_s} =  1+\epsilon\left\{ \frac{1}{2}+ \right. \\ \left. k_0\;\frac{2(\omega-k_0)\cos(\omega t)-(4 k_0+\omega)\sin(\omega t)}{\omega^2+4 k_0^2}\right\}
	\label{eq:sigmaeps}
\end{multline}
Integration of Eq.~\eqref{eq:w} with the latter approximated expression of $\sigma(t)$ gives for the average power 
\begin{equation} \label{workeasy}
P =  \frac{\omega}{2} k_0 T_c \epsilon^2 \frac{\omega-k_0}{\omega^2+4k_0^2}.
\end{equation}
Such a formula is consistent with the observation of a critical frequency separating a regime (at low frequency) where the model produces work, i.e.  $P<0$, and a regime (at high frequency) where it  adsorbs work, i.e. $P>0$: in this small $\epsilon$ limit the critical frequency is $\omega^* =  k_0$. Efficiency is more complicated to get, since it requires integrating the heat on the heat-adsorbing part of the cycle. In the next subsection we do the calculation of heat and efficiency, again in the linear regime, following a more powerful approach,
i.e. recalling the study of Onsager coefficients done in~\cite{brandner2015} and discussing the  possible optimisation strategies for the passive engine with the chosen protocols.

\subsection{Linear irreversible thermodynamics}
\label{sec:linear}

In order to exploit general results obtained in~\cite{brandner2015}, we consider  here the following choices of
 the parameter time-dependence:
 \begin{subequations}
\begin{align} 	\label{protocols}
k(t) &= k_0 + \epsilon_k \gamma_w(t), \\
T(t) &= \frac{T_c T_h}{T_h- \epsilon_T\gamma_q(t)} \approx T_c+\epsilon_T \gamma_q(t), \label{prot2}
\end{align}
\end{subequations}
with the cold temperature $T_c$, the hot temperature $T_h = T_c +\epsilon_T$ and $\gamma_w(t),\gamma_q(t)$ 
two adimensional periodic functions with period $t_{cycle}$ oscillating the first between $+1$ and $-1$ and the second between $0$ and $1$.
The new temperature protocols then oscillates with the same period between $T_c$ and $T_c+\epsilon_T$.
With such a protocol one may easily see that the weighting function for adsorbed heat, needed in Eq.~\eqref{eq:qh}, according to the recipe in Eq.~\eqref{seif}, is $w_{ads}(t) = \gamma_q(t)$.
Here, we adopt the choice $\gamma_w(t)=\sin(\omega t)$ and $\gamma_q=\frac{1}{2}[1-\cos(\omega t + \phi)]$. With such a choice for small $\epsilon$ the protocol, for $\phi=0$ is 
identical to the protocol discussed in the previous section. The advantage of form~\eqref{prot2} is the possibility of inheriting all the results presented in~\cite{brandner2015} where the linear thermodynamics study of the same model has been discussed in wide generality.

Linear thermodynamics~\cite{callen} is a framework where there are thermodynamic fluxes $J_w$ and  $J_q$, proportional to power and rate of adsorbed heat respectively, and conjugate thermodynamic forces $F_w$ and $F_q$, proportional to maximal variations of stiffness and temperature ($\epsilon_k$ and $\epsilon_T$) respectively. More precisely, one sets 
\begin{align}
J_w = \frac{P}{T_c F_w}\\
J_q = \frac{Q_h}{t_{cycle}}\\
F_w = 2 \frac{ \epsilon_k}{k_0}\\
F_q = \frac{1}{T_c}-\frac{1}{T_c+\epsilon_T} \approx \frac{\epsilon_T}{T_c^2}
\end{align}
for the fluxes and the forces respectively. We stress that in our
definitions work and power are positive when adsorbed, so that $J_w$
has the same sign of $P$, which is different from the definition
in~\cite{brandner2015}. When forces are small, $F_w \ll 1$ and $F_q
\ll 1$ it is possible to write linear relations between fluxes and
forces, through the introduction of so-called Onsager coefficients
$L_{\alpha \beta}$ with $\alpha$ and $\beta$ indices that take the
value $w$ or $q$:
\begin{align} \label{linear}
J_w &= L_{ww} F_w + L_{wq}F_q + \mathcal{O}(F^2),\\
J_q &= L_{qw} F_w + L_{qq}F_q + \mathcal{O}(F^2).
\end{align}
This immediately gives an expression for power, heat  and efficiency:
\begin{align}
P&=T_c F_w (L_{ww} F_w + L_{wq} F_q), \\
\frac{Q_h}{t_{cycle}} &=L_{qw}F_w + L_{qq}F_q, \\
\eta &= -\frac{T_c F_w(L_{ww} F_w + L_{wq}F_q)}{L_{qw}F_w + L_{qq}F_q}
\end{align}
(the reader should remember that the machine does useful work when $P<0$ and $\eta>0$).

The coefficients are given by Eqs. (72) of~\cite{brandner2015} which we rewrite with our notation:
\begin{align} \label{onsager}
L_{\alpha \beta} = -\frac{2 T_c^2 \xi_\alpha \xi_\beta}{t_{cycle}}\int_{0}^{t_{cycle}} dt [&\dot\gamma_{\alpha}(t) \gamma_\beta(t)- \\& \dot\gamma_\alpha(t)\Gamma_{\alpha \beta}(t))] \nonumber\\
\Gamma_{\alpha\beta}(t) = \int_0^\infty d\tau \dot\gamma_\beta(t-\tau)e^{-2 k_0 \tau}&\\
\xi_w=\frac{1}{4T_c} \;\;\;\; \xi_q = -\frac{1}{2} &.
\end{align}

For the present case (i.e. chosen harmonic potential and chosen temporal protocols), direct calculations give
\begin{subequations} \label{finalonsager}
\begin{align}
L_{ww}(k_0,\omega)&= \frac{k_0 \omega^2}{8(4 k_0^2 + \omega^2)}\\
L_{qq}(k_0,\omega)&= \frac{k_0 \omega^2 T_c^2}{8(4 k_0^2+\omega^2)}\\
L_{wq}(k_0,T_c,\omega,\phi)&= -\frac{k_0 \omega T_c(2 k_0 \cos(\phi)+\omega \sin(\phi))}{8(4 k_0^2 + \omega^2)}\\
L_{qw}(k_0,T_c,\omega,\phi)&= -L_{wq}(k_0,T_c,\omega,-\phi).
\end{align}
\end{subequations}
The positivity of $L_{ww}$ confirms that, in the absence of a
temperature variation, the work is always positive, i.e. it is always
adsorbed.  The relation between the off-diagonal coefficients $L_{wq}$
and $L_{qw}$ are consistent with reciprocity which is expected from
the assumption of underlying time-reversible dynamics (when in the
absence of thermodynamic forces) which here takes the form
$L_{wq}[k(t),T(t)]=L_{qw}[k(-t),T(-t)]$~\cite{brandner2015}. 

From the expressions~\eqref{finalonsager} one gets the expressions for power, heat and efficiency as functions of the model's parameters. 
\begin{subequations}
\begin{align}
P &=\frac{\omega \epsilon_k}{4 k_0}\;\frac{2\epsilon_k \omega T_c - \epsilon_Tk_0 f_+(\phi,k_0,\omega)}{4 k_0^2+\omega^2} \label{fullpower}\\
\frac{Q_h}{t_{cycle}} &=\frac{\omega}{8} \;\frac{\epsilon_T k_0 \omega + 2 \epsilon_k T_c f_-(\phi,k_0,\omega)}{4 k_0^2+\omega^2}\\
\eta &=2\frac{\epsilon_k}{k_0}\;\frac{-2 \frac{\epsilon_k}{k_0} \omega +\frac{\epsilon_T}{T_c} f_+(\phi,k_0,\omega)}{\frac{\epsilon_T}{T_c} \omega + 2 \frac{\epsilon_k}{k_0} f_-(\phi,k_0,\omega)},
\end{align}
   \end{subequations}
where we have introduced the two phase-dependent frequencies $f_\pm(\phi,k_0,\omega)=2k_0\cos(\phi)\pm\omega\sin(\phi)$.
Note that the efficiency is always lower than the Carnot efficiency which is reached when $\omega \to 0$:
\begin{equation}
\eta(\omega > 0) \le \eta(\omega=0)=   \frac{\epsilon_T}{T_c}\approx 1-\frac{T_c}{T_h}= \eta_c .
\end{equation}
Power is negative (i.e. the machine produces work) only in a range of (non-negative) frequencies, at given $\phi$, defined by
\begin{equation}
\frac{\omega}{k_0} <r(\phi)=\eta_c  \frac{2 \cos(\phi)}{2\frac{\epsilon_k}{k_0}-\eta_c\sin(\phi)},
\end{equation}
which implies that valid frequencies can be found only for ranges of $\phi$ such that $r(\phi) \ge 0$ (such ranges depend upon $\eta_c$ and $\epsilon_k/k_0$). 

In formula~\eqref{fullpower} we find interesting the role of $\phi$, which seems to us overlooked in the literature. At constant $\omega$ one can get relevant improvement in power or efficiency by tuning $\phi$: it is sufficient to consider for instance that at $\phi=0$ the range of working frequencies is $\omega<k_0 \eta_c\frac{k_0}{\epsilon_k}$, but in general such a range extends to higher frequencies when $\phi$ is increased.

In conclusion we also report the expressions for power and efficiencies  for the
 case of proportional thermodynamic forces, i.e. $\epsilon_T/T_c=\epsilon_k/k_0=\epsilon$:
 \begin{subequations} 
\begin{align} 
P &=\frac{\omega \epsilon^2 k_0 T_c}{4 } \;\frac{2 \omega  -  f_+(\phi,k_0,\omega)}{4 k_0^2+\omega^2} \label{pow}\\
\eta &=2\epsilon\;\frac{-2  \omega + f_+(\phi,k_0,\omega)}{\omega + 2  f_-(\phi,k_0,\omega)}
\end{align}
\end{subequations}
We note that power for $\phi=0$ has the same expression as in~\eqref{workeasy}.

\subsection{Optimization of power}

\begin{figure}[h!]
	\centering
	\includegraphics[width=\columnwidth]{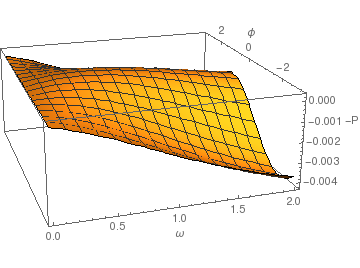}
	\caption{Delivered power $-P$ as function of $\omega$ and $\phi$, with $k_0=T_c=1$ and $\epsilon_T=\epsilon_k=0.1$.} .
	\label{fig:surface}
\end{figure}

In Fig.~\ref{fig:surface} we show the surface $-P$ as function of $\omega,\phi$, with given $k_0=T_c=1$ and $\epsilon_T=\epsilon_k=0.1$.
As a general feature, the surface has  a positive part in the low $\omega$ and low $|\phi|$ region.

Now we find the optimal phase and frequency  to get the maximum delivered power $-P$ at given $\epsilon_T$, $\epsilon_k$, $k_0$ and $T_c$. This maximum is obtained by imposing the simultaneous condition $\partial_\phi P=0$ and $\partial_\omega P=0$, and excluding solutions with $\omega \le 0$. The result of the procedure is the following formula for the optimal values $\omega^*$ and $\phi^*$ and the corresponding values of power and efficiency:
\begin{subequations} \label{optimal}
\begin{align}
\omega^*&=2 k_0\frac{\epsilon_T k_0}{\sqrt{(4 \epsilon_k T_c)^2-(\epsilon_T k_0)^2}} \label{optfr}\\
\phi^*&=\arctan\left(\frac{\omega^*}{2 k_0}\right) \label{optlag}\\
-P(\phi^*,\omega^*) &= \frac{k_0 \epsilon_T^2}{32 T_c} \\
\eta(\phi^*,\omega^*)&= \frac{\epsilon_T}{2 T_c} \approx 1-\sqrt{\frac{T_c}{T_h}}
\end{align}
\end{subequations}
Several comments are in order after looking at those formula. First of
all we notice that the optimal frequency exists only if
$\epsilon_T/T_c < 4 \epsilon_k /k_0$. Second, we confirm the interesting role of $\phi$ which
must be tuned consistently to achieve maximum power. Finally we
underline that the efficiency at maximum power is given by the
Curzon-Ahlborn formula (approximated for small $\epsilon_T$)~\cite{curzon1975}.

\section{The active heat engine}

In this Section we analyze how the previous results obtained for the
passive engine model can be exploited to get an optimal active heat
engine. In the first subsection - for the purpose of a knowledge of
all possibilities - we discuss what can be done using the dynamical UCNA
approximation (small $\tau_a$) elaborated in Section~\ref{sec:fox}
(and frequently used in the literature for problems with constant
parameters): such an approximation is useful to get a first idea of
when an active machine can do useful work, however it is not obvious
how it can be optimized. In the second subsection on the contrary we
discuss the result of the exact equivalence between the active model
and a passive one, with temperature obeying Eq.~\eqref{mappingaoup},
exploiting the optimisation strategies of the passive model.

\subsection{The small $\tau_a$ limit and the role of the active temperature in the dynamical UCNA approximation}

As discussed in Section~\ref{sec:fox} the dynamical UCNA approximation obtained in a weak active regime ($\tau_a \to 0$) constitutes an alternative mapping of an active AOUP system into a passive one, with respectively active stiffness $k_a(t)$ and temperature $T_a(t)$ given by Eqs.~\eqref{mappingfox}.
The fact that, in the dynamical UCNA approximation, the active temperature is spontaneously time-dependent even when the characteristic energy, dictated by the active speed, $v_0^2$ is constant, leads us to argue that it is in principle possible that a thermic machine is at work by modulating in time $k(t)$ only: this would be a remarkable results, in view of the fact that for passive particles it is forbidden (see Sec. IV.A) and that it would be a great advantage for experiments, where modulating $v_0$ in time can be complicate. 

To test this hypothesis we plug our simple protocol $k(t)$ (Eq.~\eqref{protocols_easy}) in the equation for the variance, obtaining $T_a(t)$. Then, similarly to what we did with the passive heat engine, we look for a formal solution of $\sigma(t)$ (by replacing $k$ with $k_a$, $T$ with $T_a$ in expression~\eqref{formalsigma}). 
Given the difficulty to write explicitly this form, we move, as usual, to the linear response regime and to a numerical approach.

\begin{figure}[h!]
\centering
\includegraphics[width=0.9\columnwidth]{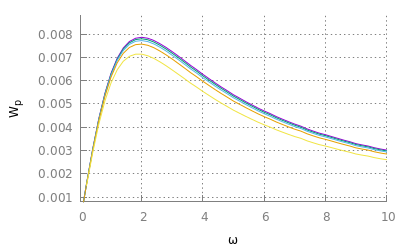} 
\includegraphics[width=0.8\columnwidth]{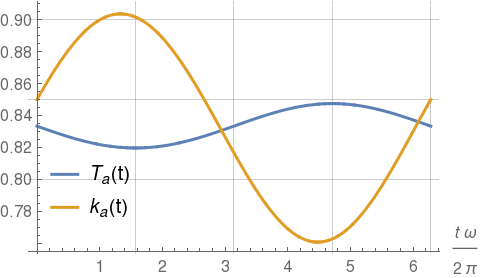} 
\includegraphics[width=0.9\columnwidth]{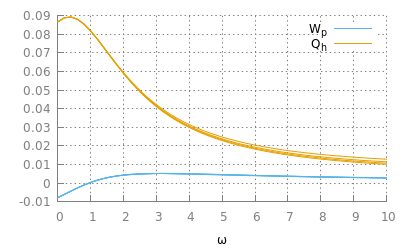} 
\includegraphics[width=0.8\columnwidth]{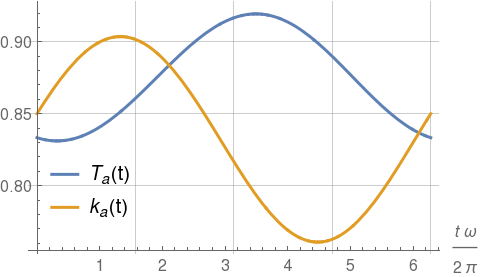}
\caption{Study of thermodynamics for the AOUP model at small $\tau$. A: work per cycle $W_p$ as a function of $\omega$ in the case of constant self-propulsion speed $v_0$: no work is produced for $\tau_a=0.01$ (higher curve)$-0.1$ (lower curve). B: Active stiffness and temperature (in the case with $\tau_a=0.1, \omega=1 $) when $v_0^2 \tau_a=1$ is constant: maxima of $T_a(t)$ are always in phase with minima of $k_a(t)$, failing to meet a working machine condition. C: Work $W_p$ and adsorbed heat $Q_h$ per cycle when $v_0(t)$ is time-dependent. D: Active stiffness and temperature (with $\tau_a=0.2, \omega=1$) in the time-dependent $v_0(t)$ case. The time modulation of $v_0$ in C,D occurs with parameters $\tau_a u^2=1,\epsilon_u=0.1 u^2 $. In all plots $k_0=1$ and $\epsilon_k=0.1$.} \label{fig:fox}
\end{figure}

The numerical integration betrays our expectations showing that work in a cycle is positive at any frequency $\omega$ (Fig.~\ref{fig:fox})A. 
This is furthermore verified by the small perturbation in $\epsilon$. We proceed as in Section~\ref{sec:qualitative}, expanding $\sigma(t)=\sigma_0(t) + \epsilon \sigma_1(t)$, and computing an approximated expression for $k_a$ and $T_a$ to be inserted in~\eqref{eq:sigma}. 
\begin{subequations} \label{eq:Taka_eps}
\begin{align}
k_a &= \frac{k_0}{\Gamma_0} + 
\epsilon \frac{k_0}{\Gamma_0^2} [\Gamma_0 \omega \tau_a  \cos(\omega t) + \sin(\omega t)] \\
T_a &= \frac{v_0^2 \tau_a}{\Gamma_0^2} \bigg[ 1 - \epsilon  \frac{2 \tau_a k_0}{\Gamma_0}  \sin(\omega t) \bigg]
\end{align}	
\end{subequations}
We have used $\Gamma_0=1+\tau_a k_0$.

The related work $W_p = \epsilon^2 (v_0^2 \tau_a ) \frac{k_0}{2\Gamma_0} \: \: \frac{\omega^2}{4 k_0^2 + (\Gamma_0 \omega)^2}$ is positive and does not cross 0 for any frequency value $\omega >0$. Note that in the passive limit we recover the expression for the power~\eqref{workeasy0}.

The fact that the pure modulation of $k(t)$, i.e. keeping $v_0$ constant, does not produce a working machine in the small $\tau$ limit, can be understood on a more general ground, i.e. independently of the small $\epsilon$ limit and of the choice of the protocol $k(t)$.
In fact, following the qualitative discussion given in section~IV, we suggest that the form of $T_a(t)$ with $v_0$ constant does not meet the requirement of a working Stirling  engine: the expansions ($\dot k_a <0$)  are not in phase with the maximum temperature $T_a$ (See Fig.~\ref{fig:fox}B as an example).
Given the positivity of $k$, it is straigthforward to see that the following constraints are never satisfied at the same time:
\begin{equation}
	\begin{cases}
		\dot{T}_a = 0 \quad\ ;\quad\ \ddot{T}_a >0\\
		\dot{k}_a <0
	\end{cases}
	\label{eq:con}
\end{equation}
The presence of a lag between stiffness and temperature, such that the temperature maximum is in the expansion phase of the confining potential is decisive in the realisation of a working engine, similarly to the passive case. We need to let $v_0(t)$ vary in time, in order to force $T_a(t)$ to take the required form.\\

\begin{figure}[h!]
	\centering
	\includegraphics[width=\columnwidth]{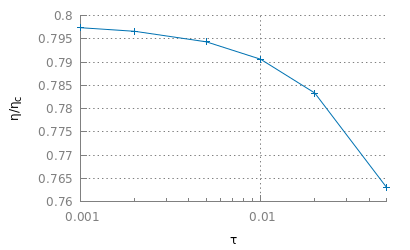}
	\caption{Efficiency - rescaled by the Carnot efficiency in the passive limit, $\eta_c=\epsilon_u/u^2=0.1$ - in small $\tau$ AOUP model when $v_0(t)$ follows protocol in Eq.~\eqref{v0ucna}, inspired by dynamical UCNA approximation. The efficiency decreases with activity. Parameters: $\omega=0.1,\epsilon_k=0.1,k_0=1$, $u^2 =1/\tau_a$ and $\epsilon_u=0.1 u^2$} \label{fig:foxefficiency}
\end{figure}

For small $\tau_a$ note that $k_a \to k$ and $T_a \to T_D=v_0^2 \tau_a$, so it is natural to propose a $v_0$ which resembles the passive temperature in ~\eqref{protocols_easy}. We take
\begin{equation} \label{v0ucna}
	v_0^2(t)= u^2 + {\epsilon_u} \frac{1 - \cos(\omega t)}{2}
\end{equation}
This intuition is qualitatively right: the active AOUP model with a
time-dependent typical velocity is able to produce work, as we can see
in Fig.~\ref{fig:fox}C. Along with the numerical result, it is
possible to repeat our linearization strategy for $\sigma(t)$ and show
that the power, obtained with the correct approximation for $T_a$, is
$P = \epsilon^2 (v_0^2 \tau_a ) \frac{k_0}{2\Gamma_0} \: \:
\frac{\omega(\omega - k_0)}{4 k_0^2 + (\Gamma_0 \omega)^2}$. We stress
out a regime change for $\omega=k_0$ and the agreement with expression
~\eqref{workeasy} in the passive limit. In order to evaluate the
efficiency of this working machine, in particular its behavior with
$\tau_a$, we resort to numerical integration of adsorbed
heat and
work. For adosrbed heat we use definition~\eqref{seif} for $w_{ads}(t)$, using $T_D(t)=\tau_a v_0^2(t)$ in place of $T(t)$. The results for efficiency are represented in
Fig.\ref{fig:foxefficiency}: we emphasize that it is maximum in the
quasi-static and in the passive limits. The effect of self-propulsion, within this approach,
seems to decrease efficiency, but this is basically due to the fact that the chosen protocol is not sensitive to $\tau_a$ and $k(t)$, while the real effective temperature $T_{eff}$ is. Changing $\tau$ without adapting the protocol degrades the efficiency.

In the next section we explore the aforementioned passive-active equivalence, which gives the possibility to adjust the protocol when $\tau_a$ is varied, in order to control power and efficiency of the engine.

\subsection{Optimisation by passive equivalence}


The idea of exploiting passive-active equivalence is the
following. Whatever is the particular optimisation procedure applied
to the passive model, one gets optimal passive protocols $k^*(t)$ and
$T^*(t)$. At that point the "mapping equation",
Eq.~\eqref{mappingaoup}, can be used to derive the corresponding
protocols for the active models: such protocols will give exactly the
same power and the same efficiency and therefore will be optimal in
that particular protocols' set. Note that, if the passive engine is
optimised in the family of protocols $k(t),T(t)$ given by
Equations~\eqref{protocols}, with parameters
$k_0,T_c,\epsilon_k,\epsilon_T,\omega,\phi$, the family of protocols
which is spanned in the optimisation procedure is given by the same
$k(t)$ and a function $v_0(t)$ which satisfies Eq.~\eqref{mappingaoup}
with $T_{eff}=T(t)$. Putting Eqs.~\eqref{protocols} into
Eq.~\eqref{mappingaoup}, we get the corresponding family of protocols
for $v_0(t)$
\begin{subequations} \label{v0q}
\begin{align} 
  &v_0^2(t) \tau_a = T(t)+\tau_a k(t) T(t)+\\
  &\frac{3}{4}\omega \tau_a\epsilon_T \sin(\omega t+\phi)+\\
  &\frac{\omega \tau_a^2}{2} \left[\epsilon_k \cos(\omega t) T(t)+\frac{\epsilon_T}{2}\sin(\omega t+\phi)k(t)\right]+\\
  &\frac{\omega^2\tau_a^2\epsilon_T }{4}\cos(\omega t+\phi),
\end{align}
\end{subequations}
which is parametrized by $k_0,T_c,\epsilon_k,\epsilon_T,\omega,\phi$
and also $\tau_a$.

Summing up, if $\tau_a$ and $k(t)$ are imposed by the experiment, and
one looks for an optimal $v_0(t)$, the task is relatively easy,
i.e. one may 1) choose arbitrary values for $T_c$ and small $\epsilon_T$ (see below) and then 2) directly find the optimal $\omega^*$ and $\phi^*$ for the
passive problem (in the family of sinusoidal passive protocols given
by Eqs.~\eqref{protocols}), i.e. formula in Eqs.~\eqref{optimal} and
finally 3) use formula~\eqref{v0q}, to get the corresponding active optimal
protocol for $v_0(t)$ which guarantees the maximum possible power and
a corresponding Curzon-Ahlborn efficiency, {\em whatever is the value of $\tau_a$}. Given that $\epsilon_T/T_c$ must be small and therefore it is not really a free number (reasonable values are $0.1$ or smaller), some freedom remains in choosing $T_c$, which can be exploited in two ways: 1) one may set the desired optimal frequency $\omega^*$ (based upon possible experimental requirements) and then invert equation~\eqref{optfr} to get the corresponding $T_c$ or, in alternative, 2) observe that Eq.~\eqref{mappingaoup} is invariant for common rescaling of $T_{eff}$ (and therefore $T_c$) and $v_0^2$, that is one may meet any experimental upper or lower limit for $v_0^2$ by accordingly rescaling $T_c$.

\begin{figure}[t]
  \includegraphics[width=\columnwidth]{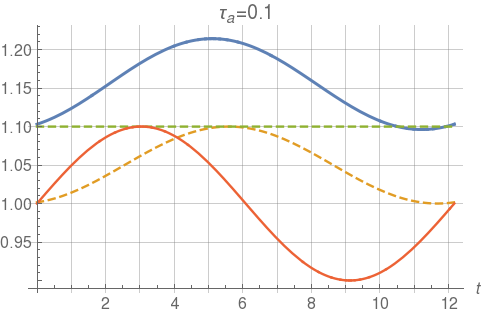}
  \includegraphics[width=0.97\columnwidth]{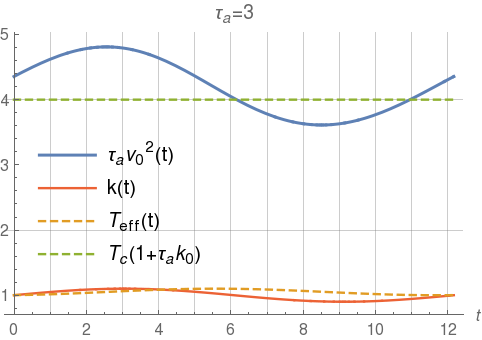}
	\caption{Examples of the optimal protocol $\tau_a v_0^2(t)$ for an active engine to achieve maximum power
          when $k_0=T_c=1$ and $\epsilon_T=\epsilon_k=0.1$ (which,
          according to formula~\eqref{optimal}, give $\omega^* \approx
          0.52$ and $\phi^* \approx 0.25$) and $\tau_a=0.1$. Two values are considered: $\tau_a=0.1$ and $\tau_a=3$. We underline that in both cases the engine gives the same maximum power $\approx 3 \; 10^{-4}$ and the same Curzon-Ahlborn efficiency $\approx 0.05$. The blue curve is $\tau_a
          v_0^2(t)$ and the red curve is $k(t)$. For reference we also
          put $T(t)=T_{eff}(t)=T_c+\epsilon_T\frac{1}{2}[1-\cos(\omega^*t+\phi^*)]$
          (yellow dashed curve), and the constant $T_c(1+\tau_a
          k_0)$ (green dashed curve) which is the approximation of
          Eq.~\eqref{v0q} at $0$ order in
          $\omega,\epsilon_k,\epsilon_T$.
	\label{fig:5}}
\end{figure}

In Figure~\ref{fig:5} we show the optimal protocols for a given choice
of $T_c,k_0,\epsilon_T,\epsilon_k$. As anticipated, there is an
important difference between the optimal protocol for $\tau_a
v_0^2(t)$ and $T_{eff}(t)$. We underline that, following this
strategy, if one spans a range of $\tau_a$ - keeping the same $k(t)$ -
the optimal effective temperature $T_{eff}(t)$ is not changed: what
is changed is the corresponding protocol for $v_0(t)$; if one follows
it, whatever is the value of $\tau_a$, the power and the efficiency of
the engine will always be the same. Also for this reason it is useless
to show a plot with efficiencies as a function of $\tau_a$. The
constancy of power and efficiency as a function of $\tau_a$ already
demonstrates the superiority of this approach with respect to other
approaches not informed with the correct formula for $T_{eff}$ (for
instance the one of previous Section, where the efficiency decays with
$\tau_a$, see Figure~\ref{fig:foxefficiency}).


  The situation is more complicate if $\tau_a$ and $v_0(t)$ are imposed by the experiment and one wants to look
  for the optimal $k(t)$. In such a case, a possible strategy is
  to use Eq.~\eqref{mappingaoup} to get a functional constraint
  between $k(t)$ and $T_{eff}(t)$; thereafter, one needs to solve the
  passive problem with a variation of the coupled protocols
  $k(t),T_{eff}(t)$ with the given constraint.

Strategies of passive-to-active equivalence are substantially simplified  if
very slow transformations are considered, i.e. in the limit of large
period $t_{cycle}$, more precisely by taking $\tau_a/t_{cycle} \ll
1$. In this limit Eq.~\eqref{mappingaoup} is considerably simpler, as
it reduces to the  identity (valid for any magnitude of forces $\epsilon_k$ and $\epsilon_T$)
\begin{equation}
T_{eff}(t) =\frac{\tau_a v_0^2(t)}{1+k(t) \tau_a}. \label{mappingqs}
\end{equation}
As mentioned, in this limit $T_{eff}(t)$ is still different from
$T_a(t)$, even at first order in $\tau_a$, see Eq.~\eqref{ineq}. We underline  that the same problem discussed above (see discussion above Eqs.~\eqref{eq:con}) occurs for the expression of $T_{eff}(t)$ in formula~\eqref{mappingqs}: if $v_0(t)$ is taken constant, the resulting effective temperature is always in opposition of phase with $k(t)$ (ie. maxima of $k$ correspond to minima of $T_{eff}$ and viceversa). Several empirical attempts, by numerical integration of $W_p$ for a wide range of choices of all the parameters convinced us that such a situation always leads to $W_p \ge 0$ ie. a machine that does not produce work. We recall that this is rigorously proven in the linear forcing regime (see Section IV.B and formula~\eqref{fullpower}, the opposition of phase between $k(t)$ and $T(t)$ corresponds to $\phi=-\pi/2$).

Equation~\eqref{mappingqs} gives an extimate of the
maximum efficiency (to be attained in the $\omega \to 0$ limit, that is at vanishing power), i.e.
\begin{equation}
\eta_c = 1-\min\left\{ \frac{v_0^2(t)}{1+\tau_a k(t)} \right\} \max\left\{ \frac{v_0^2(t)}{1+\tau_a k(t)} \right\}^{-1} \label{acteff}
\end{equation}
which is a striking evidence of the non-trivial relation between the
two thermodynamic forces (for temperature and volume forces) in shaping the
efficiency of active heat engines. 


Equation~\eqref{mappingqs} can be also used to find the shape of $v_0(t)$ to get any desired efficiency $\eta_c$ (at vanishing power, $\omega \to 0$). This is achieved by imposing that $\eta_c=\epsilon_T/(T_c+\epsilon_T)$ and recalling the non-linear expression for $T(t)$, see~\eqref{prot2}. Then we obtain
\begin{equation} \label{alleff}
v_0^2(t)=\frac{T_c}{\tau_a}\frac{1+\tau_a k(t)}{1-\eta_c \gamma_q(t)} \xrightarrow{\tau_a\rightarrow\infty} \frac{T_c k(t)}{1-\eta_c \gamma_q(t)}.
\end{equation}
We warn, however, that Eq.~\eqref{mappingqs} only guarantees that the
passive model with temperature $T_{eff}(t)$ gives the same evolution
for $\sigma(t)$ and therefore produces/adsorbs the same work, but is
not necessarily a heat engine. The positivity of work production (which in our notation corresponds to $W_p<0$) depends upon the phase shift between $T_{eff}(t)$ and $k(t)$. Therefore, in Eq.~\eqref{alleff} one needs to put the proper $k(t)$ and $\gamma_q(t)$, i.e. the correct choices of $\omega$, $k_0$ and $\phi$ : the desired efficiency is reached, provided that the machine does useful work. We know however that, for small $\epsilon_k$ and $\epsilon_T$ (as demonstrated by Eq.~\eqref{pow} in the $\omega \to 0$ limit), such a condition is  satisfied for $\phi \in (-\pi/2, \pi/2)$.

\section{Conclusions}

A well defined thermostat temperature is a crucial ingredient for definitions of basic thermodynamic tools (e.g. adsorbed heat and efficiency) as well as to transfer known results valid for thermal systems: for the lack of such a well defined temperature, active heat engines elude intuition and expectation in stochastic thermodynamics. Here, building upon an important observation made in~\cite{holubec2020}, we have shown an example where such a temperature can be defined and gives important advantages, useful also in experiments.

The effective temperature for this particular model satisfies Eq.~\eqref{mappingaoup} and is different from all other temperatures based upon particular - usually static - configurations (e.g. $T_D$ related to unconfined diffusion, $T_{var}$ related to equilibrium steady states, $T_a$ related to the small $\tau_a$ limit, etc.). It represents, exactly, the thermostat of an equivalent passive model which gives - in the presence of the same external harmonic potential - the same position variance and therefore the same power and the same total heat exchanged. An observation about the most correct definition of adsorbed heat (see discussion at the end of Section II) suggests that the equivalence noted in~\cite{holubec2020} cannot be extended to efficiency. Strictly speaking, efficiency of a model without a temperature is not defined at all because it is not evident how to discriminate, properly, between heat adsorbed and heat dissipated. We have bypassed this conceptual point, by considering the efficiency of the equivalent passive model. Since it is designed to give Carnot efficiency in the quasi-static limit, whatever the values of other parameters, one can use it as a proper figure of merit for the purpose of evaluating the performance of the machine.

The active-passive equivalence, Eq.~\eqref{mappingaoup}, which contains time-derivatives of $T(t)$ and $k(t)$, suggests to study the optimisation of a passive model with smooth protocols, that is different from what usually done with piecewise linear modulations (for Carnot-like or Stirling-like engines). Therefore we have extended previous studies to a family of smooth protocols where the lag (between temperature and stiffness modulation) is varied to improve efficiency. In the linear approximation of fluxes we have found the optimal frequency and phase lag (Eqs.~\eqref{optfr}-\eqref{optlag}) that produces maximum power output (and correspondingly Curzon-Ahlborn efficiency, roughly half of Carnot efficiency), a result which is readily translated to active engines through Eq.~\eqref{mappingaoup}. This equivalence equation also immediately gives the Carnot efficiency of an active engine, see Eq.~\eqref{acteff}, which is valid for any activity time $\tau_a$ (i.e. also far from the passive limit) and any amplitude of the protocols (i.e. also far from the linear regime), but of course can be attained only in the quasi-static limit, that is at vanishing power.

Future investigations concern the possibility of extending, through suitable approximations, the results of our study to non-harmonic potentials~\cite{holubec2020}, as well as to other active particle models. It is also interesting to consider fluctuations of the relevant quantities, such as power or efficiency, which constitute an important ingredient of microscopic engines. Finally, a promising direction of research is to  consider bunches of active particles with interactions, in order to probe the effect of collective behavior  on the performance of such kinds of heat machines.

\begin{acknowledgments}
The Authors acknowledge useful discussions with Andrea Baldassarri concerning the definition of efficiency. They also thank Lorenzo Caprini, Umberto Marini Bettolo Marconi and Alessandro Sarracino for several useful comments. AP acknowledges financial support from MIUR through the PRIN 2017 grant number 201798CZLJ and from Regione Lazio through the Grant “Progetti Gruppi di Ricerca” N. 85-2017-15257.
\end{acknowledgments} 

\bibliography{biblio}

\begin{thebibliography}{47}%
\makeatletter
\providecommand \@ifxundefined [1]{%
 \@ifx{#1\undefined}
}%
\providecommand \@ifnum [1]{%
 \ifnum #1\expandafter \@firstoftwo
 \else \expandafter \@secondoftwo
 \fi
}%
\providecommand \@ifx [1]{%
 \ifx #1\expandafter \@firstoftwo
 \else \expandafter \@secondoftwo
 \fi
}%
\providecommand \natexlab [1]{#1}%
\providecommand \enquote  [1]{``#1''}%
\providecommand \bibnamefont  [1]{#1}%
\providecommand \bibfnamefont [1]{#1}%
\providecommand \citenamefont [1]{#1}%
\providecommand \href@noop [0]{\@secondoftwo}%
\providecommand \href [0]{\begingroup \@sanitize@url \@href}%
\providecommand \@href[1]{\@@startlink{#1}\@@href}%
\providecommand \@@href[1]{\endgroup#1\@@endlink}%
\providecommand \@sanitize@url [0]{\catcode `\\12\catcode `\$12\catcode
  `\&12\catcode `\#12\catcode `\^12\catcode `\_12\catcode `\%12\relax}%
\providecommand \@@startlink[1]{}%
\providecommand \@@endlink[0]{}%
\providecommand \url  [0]{\begingroup\@sanitize@url \@url }%
\providecommand \@url [1]{\endgroup\@href {#1}{\urlprefix }}%
\providecommand \urlprefix  [0]{URL }%
\providecommand \Eprint [0]{\href }%
\providecommand \doibase [0]{http://dx.doi.org/}%
\providecommand \selectlanguage [0]{\@gobble}%
\providecommand \bibinfo  [0]{\@secondoftwo}%
\providecommand \bibfield  [0]{\@secondoftwo}%
\providecommand \translation [1]{[#1]}%
\providecommand \BibitemOpen [0]{}%
\providecommand \bibitemStop [0]{}%
\providecommand \bibitemNoStop [0]{.\EOS\space}%
\providecommand \EOS [0]{\spacefactor3000\relax}%
\providecommand \BibitemShut  [1]{\csname bibitem#1\endcsname}%
\let\auto@bib@innerbib\@empty
\bibitem [{\citenamefont {Feynamn}(1960)}]{feynman59}%
  \BibitemOpen
  \bibfield  {author} {\bibinfo {author} {\bibfnamefont {R.~P.}\ \bibnamefont
  {Feynamn}},\ }\href@noop {} {\bibfield  {journal} {\bibinfo  {journal} {Eng.
  Sci.}\ }\textbf {\bibinfo {volume} {23}},\ \bibinfo {pages} {22} (\bibinfo
  {year} {1960})}\BibitemShut {NoStop}%
\bibitem [{\citenamefont {Blickle}\ and\ \citenamefont
  {Bechinger}(2012)}]{blickle12}%
  \BibitemOpen
  \bibfield  {author} {\bibinfo {author} {\bibfnamefont {V.}~\bibnamefont
  {Blickle}}\ and\ \bibinfo {author} {\bibfnamefont {C.}~\bibnamefont
  {Bechinger}},\ }\href@noop {} {\bibfield  {journal} {\bibinfo  {journal}
  {Nat. Phys.}\ }\textbf {\bibinfo {volume} {8}},\ \bibinfo {pages} {143}
  (\bibinfo {year} {2012})}\BibitemShut {NoStop}%
\bibitem [{\citenamefont {Ro{\ss}nagel}\ \emph {et~al.}(2014)\citenamefont
  {Ro{\ss}nagel}, \citenamefont {Abah}, \citenamefont {Schmidt-Kaler},
  \citenamefont {Singer},\ and\ \citenamefont {Lutz}}]{rossnagel14}%
  \BibitemOpen
  \bibfield  {author} {\bibinfo {author} {\bibfnamefont {J.}~\bibnamefont
  {Ro{\ss}nagel}}, \bibinfo {author} {\bibfnamefont {O.}~\bibnamefont {Abah}},
  \bibinfo {author} {\bibfnamefont {F.}~\bibnamefont {Schmidt-Kaler}}, \bibinfo
  {author} {\bibfnamefont {K.}~\bibnamefont {Singer}}, \ and\ \bibinfo {author}
  {\bibfnamefont {E.}~\bibnamefont {Lutz}},\ }\href@noop {} {\bibfield
  {journal} {\bibinfo  {journal} {Phys. Rev. Lett.}\ }\textbf {\bibinfo
  {volume} {112}},\ \bibinfo {pages} {030602} (\bibinfo {year}
  {2014})}\BibitemShut {NoStop}%
\bibitem [{\citenamefont {Martinez}\ \emph
  {et~al.}(2015{\natexlab{a}})\citenamefont {Martinez}, \citenamefont
  {Rold\'an}, \citenamefont {Dinis}, \citenamefont {Petrov}, \citenamefont
  {Parrondo},\ and\ \citenamefont {Rica}}]{martinez15}%
  \BibitemOpen
  \bibfield  {author} {\bibinfo {author} {\bibfnamefont {I.}~\bibnamefont
  {Martinez}}, \bibinfo {author} {\bibfnamefont {E.}~\bibnamefont {Rold\'an}},
  \bibinfo {author} {\bibfnamefont {L.}~\bibnamefont {Dinis}}, \bibinfo
  {author} {\bibfnamefont {D.}~\bibnamefont {Petrov}}, \bibinfo {author}
  {\bibfnamefont {J.~M.~R.}\ \bibnamefont {Parrondo}}, \ and\ \bibinfo {author}
  {\bibfnamefont {R.~A.}\ \bibnamefont {Rica}},\ }\href@noop {} {\bibfield
  {journal} {\bibinfo  {journal} {Nat. Phys.}\ }\textbf {\bibinfo {volume}
  {12}},\ \bibinfo {pages} {67} (\bibinfo {year}
  {2015}{\natexlab{a}})}\BibitemShut {NoStop}%
\bibitem [{Note1()}]{Note1}%
  \BibitemOpen
  \bibinfo {note} {The macroscopic world has additional sources of energy,
  unfortunately - for our environment - of minor importance for the moment,
  such as those related to natural macroscopic flows, e.g. air and
  water}\BibitemShut {NoStop}%
\bibitem [{\citenamefont {Martinez}\ \emph {et~al.}(2017)\citenamefont
  {Martinez}, \citenamefont {Roldan}, \citenamefont {Dinis},\ and\
  \citenamefont {Rica}}]{martinez17}%
  \BibitemOpen
  \bibfield  {author} {\bibinfo {author} {\bibfnamefont {I.~A.}\ \bibnamefont
  {Martinez}}, \bibinfo {author} {\bibfnamefont {E.}~\bibnamefont {Roldan}},
  \bibinfo {author} {\bibfnamefont {L.}~\bibnamefont {Dinis}}, \ and\ \bibinfo
  {author} {\bibfnamefont {R.~A.}\ \bibnamefont {Rica}},\ }\href@noop {}
  {\bibfield  {journal} {\bibinfo  {journal} {Soft Matter}\ }\textbf {\bibinfo
  {volume} {13}},\ \bibinfo {pages} {22} (\bibinfo {year} {2017})}\BibitemShut
  {NoStop}%
\bibitem [{\citenamefont {Verley}\ \emph {et~al.}(2014)\citenamefont {Verley},
  \citenamefont {Willaert}, \citenamefont {den Broeck}, ,\ and\ \citenamefont
  {Esposito}}]{verley14}%
  \BibitemOpen
  \bibfield  {author} {\bibinfo {author} {\bibfnamefont {G.}~\bibnamefont
  {Verley}}, \bibinfo {author} {\bibfnamefont {T.}~\bibnamefont {Willaert}},
  \bibinfo {author} {\bibfnamefont {C.~V.}\ \bibnamefont {den Broeck}}, , \
  and\ \bibinfo {author} {\bibfnamefont {M.}~\bibnamefont {Esposito}},\
  }\href@noop {} {\bibfield  {journal} {\bibinfo  {journal} {Phys. Rev. E}\
  }\textbf {\bibinfo {volume} {90}},\ \bibinfo {pages} {052145} (\bibinfo
  {year} {2014})}\BibitemShut {NoStop}%
\bibitem [{\citenamefont {Pietzonka}\ and\ \citenamefont
  {Seifert}(2018)}]{pietzonka18}%
  \BibitemOpen
  \bibfield  {author} {\bibinfo {author} {\bibfnamefont {P.}~\bibnamefont
  {Pietzonka}}\ and\ \bibinfo {author} {\bibfnamefont {U.}~\bibnamefont
  {Seifert}},\ }\href@noop {} {\bibfield  {journal} {\bibinfo  {journal} {Phys.
  Rev. Lett.}\ }\textbf {\bibinfo {volume} {120}},\ \bibinfo {pages} {190602}
  (\bibinfo {year} {2018})}\BibitemShut {NoStop}%
\bibitem [{\citenamefont {Martinez}\ \emph
  {et~al.}(2015{\natexlab{b}})\citenamefont {Martinez}, \citenamefont
  {Rold\'an}, \citenamefont {Dinis}, \citenamefont {Petrov},\ and\
  \citenamefont {Rica}}]{martinez15b}%
  \BibitemOpen
  \bibfield  {author} {\bibinfo {author} {\bibfnamefont {I.~A.}\ \bibnamefont
  {Martinez}}, \bibinfo {author} {\bibfnamefont {E.}~\bibnamefont {Rold\'an}},
  \bibinfo {author} {\bibfnamefont {L.}~\bibnamefont {Dinis}}, \bibinfo
  {author} {\bibfnamefont {D.}~\bibnamefont {Petrov}}, \ and\ \bibinfo {author}
  {\bibfnamefont {R.~A.}\ \bibnamefont {Rica}},\ }\href@noop {} {\bibfield
  {journal} {\bibinfo  {journal} {Phys. Rev. Lett.}\ }\textbf {\bibinfo
  {volume} {114}},\ \bibinfo {pages} {120601} (\bibinfo {year}
  {2015}{\natexlab{b}})}\BibitemShut {NoStop}%
\bibitem [{\citenamefont {Ro{\ss}nagel}\ \emph {et~al.}(2016)\citenamefont
  {Ro{\ss}nagel}, \citenamefont {Dawkins}, \citenamefont {Tolazzi},
  \citenamefont {Abah}, \citenamefont {Lutz}, \citenamefont {Schmidt-Kaler},\
  and\ \citenamefont {Singer}}]{rossnagel16}%
  \BibitemOpen
  \bibfield  {author} {\bibinfo {author} {\bibfnamefont {J.}~\bibnamefont
  {Ro{\ss}nagel}}, \bibinfo {author} {\bibfnamefont {S.~T.}\ \bibnamefont
  {Dawkins}}, \bibinfo {author} {\bibfnamefont {K.~N.}\ \bibnamefont
  {Tolazzi}}, \bibinfo {author} {\bibfnamefont {O.}~\bibnamefont {Abah}},
  \bibinfo {author} {\bibfnamefont {E.}~\bibnamefont {Lutz}}, \bibinfo {author}
  {\bibfnamefont {F.}~\bibnamefont {Schmidt-Kaler}}, \ and\ \bibinfo {author}
  {\bibfnamefont {K.}~\bibnamefont {Singer}},\ }\href@noop {} {\bibfield
  {journal} {\bibinfo  {journal} {Science}\ }\textbf {\bibinfo {volume}
  {352}},\ \bibinfo {pages} {325} (\bibinfo {year} {2016})}\BibitemShut
  {NoStop}%
\bibitem [{\citenamefont {Schmiedl}\ and\ \citenamefont
  {Seifert}(2008)}]{schmiedl08}%
  \BibitemOpen
  \bibfield  {author} {\bibinfo {author} {\bibfnamefont {T.}~\bibnamefont
  {Schmiedl}}\ and\ \bibinfo {author} {\bibfnamefont {U.}~\bibnamefont
  {Seifert}},\ }\href@noop {} {\bibfield  {journal} {\bibinfo  {journal}
  {Europhys. Lett.}\ }\textbf {\bibinfo {volume} {81}},\ \bibinfo {pages}
  {20003} (\bibinfo {year} {2008})}\BibitemShut {NoStop}%
\bibitem [{\citenamefont {Cerino}\ \emph {et~al.}(2016)\citenamefont {Cerino},
  \citenamefont {Puglisi},\ and\ \citenamefont {Vulpiani}}]{cerino16}%
  \BibitemOpen
  \bibfield  {author} {\bibinfo {author} {\bibfnamefont {L.}~\bibnamefont
  {Cerino}}, \bibinfo {author} {\bibfnamefont {A.}~\bibnamefont {Puglisi}}, \
  and\ \bibinfo {author} {\bibfnamefont {A.}~\bibnamefont {Vulpiani}},\
  }\href@noop {} {\bibfield  {journal} {\bibinfo  {journal} {Phys. Rev. E}\
  }\textbf {\bibinfo {volume} {93}},\ \bibinfo {pages} {042116} (\bibinfo
  {year} {2016})}\BibitemShut {NoStop}%
\bibitem [{\citenamefont {Puglisi}\ \emph {et~al.}(2017)\citenamefont
  {Puglisi}, \citenamefont {Sarracino},\ and\ \citenamefont
  {Vulpiani}}]{puglisi2017temperature}%
  \BibitemOpen
  \bibfield  {author} {\bibinfo {author} {\bibfnamefont {A.}~\bibnamefont
  {Puglisi}}, \bibinfo {author} {\bibfnamefont {A.}~\bibnamefont {Sarracino}},
  \ and\ \bibinfo {author} {\bibfnamefont {A.}~\bibnamefont {Vulpiani}},\
  }\href@noop {} {\bibfield  {journal} {\bibinfo  {journal} {Phys. Rep.}\
  }\textbf {\bibinfo {volume} {709}},\ \bibinfo {pages} {1} (\bibinfo {year}
  {2017})}\BibitemShut {NoStop}%
\bibitem [{\citenamefont {Bechinger}\ \emph {et~al.}(2016)\citenamefont
  {Bechinger}, \citenamefont {{Di Leonardo}}, \citenamefont {L\"owen},
  \citenamefont {Reichhardt}, \citenamefont {Volpe},\ and\ \citenamefont
  {Volpe}}]{dileo16}%
  \BibitemOpen
  \bibfield  {author} {\bibinfo {author} {\bibfnamefont {C.}~\bibnamefont
  {Bechinger}}, \bibinfo {author} {\bibfnamefont {R.}~\bibnamefont {{Di
  Leonardo}}}, \bibinfo {author} {\bibfnamefont {H.}~\bibnamefont {L\"owen}},
  \bibinfo {author} {\bibfnamefont {C.}~\bibnamefont {Reichhardt}}, \bibinfo
  {author} {\bibfnamefont {G.}~\bibnamefont {Volpe}}, \ and\ \bibinfo {author}
  {\bibfnamefont {G.}~\bibnamefont {Volpe}},\ }\href@noop {} {\bibfield
  {journal} {\bibinfo  {journal} {Rev. Mod. Phys.}\ }\textbf {\bibinfo {volume}
  {88}},\ \bibinfo {pages} {045006} (\bibinfo {year} {2016})}\BibitemShut
  {NoStop}%
\bibitem [{\citenamefont {Leonardo}\ \emph {et~al.}(2010)\citenamefont
  {Leonardo}, \citenamefont {Angelani}, \citenamefont {Dell’Arciprete},
  \citenamefont {Ruocco}, \citenamefont {Iebba}, \citenamefont {Schippa},
  \citenamefont {Conte}, \citenamefont {Mecarini}, \citenamefont {Angelis},\
  and\ \citenamefont {Fabrizio}}]{dileonardo10}%
  \BibitemOpen
  \bibfield  {author} {\bibinfo {author} {\bibfnamefont {R.~D.}\ \bibnamefont
  {Leonardo}}, \bibinfo {author} {\bibfnamefont {L.}~\bibnamefont {Angelani}},
  \bibinfo {author} {\bibfnamefont {D.}~\bibnamefont {Dell’Arciprete}},
  \bibinfo {author} {\bibfnamefont {G.}~\bibnamefont {Ruocco}}, \bibinfo
  {author} {\bibfnamefont {V.}~\bibnamefont {Iebba}}, \bibinfo {author}
  {\bibfnamefont {S.}~\bibnamefont {Schippa}}, \bibinfo {author} {\bibfnamefont
  {M.}~\bibnamefont {Conte}}, \bibinfo {author} {\bibfnamefont
  {F.}~\bibnamefont {Mecarini}}, \bibinfo {author} {\bibfnamefont {F.~D.}\
  \bibnamefont {Angelis}}, \ and\ \bibinfo {author} {\bibfnamefont {E.~D.}\
  \bibnamefont {Fabrizio}},\ }\href@noop {} {\bibfield  {journal} {\bibinfo
  {journal} {Proc. Natl. Acad. Sci.}\ }\textbf {\bibinfo {volume} {107}},\
  \bibinfo {pages} {9541} (\bibinfo {year} {2010})}\BibitemShut {NoStop}%
\bibitem [{\citenamefont {Sokolov}\ \emph {et~al.}(2010)\citenamefont
  {Sokolov}, \citenamefont {Apodaca}, \citenamefont {Grzybowski},\ and\
  \citenamefont {Aranson}}]{sokolov10}%
  \BibitemOpen
  \bibfield  {author} {\bibinfo {author} {\bibfnamefont {A.}~\bibnamefont
  {Sokolov}}, \bibinfo {author} {\bibfnamefont {M.~M.}\ \bibnamefont
  {Apodaca}}, \bibinfo {author} {\bibfnamefont {B.~A.}\ \bibnamefont
  {Grzybowski}}, \ and\ \bibinfo {author} {\bibfnamefont {I.~S.}\ \bibnamefont
  {Aranson}},\ }\href@noop {} {\bibfield  {journal} {\bibinfo  {journal} {Proc.
  Natl. Acad. Sci.}\ }\textbf {\bibinfo {volume} {107}},\ \bibinfo {pages}
  {969} (\bibinfo {year} {2010})}\BibitemShut {NoStop}%
\bibitem [{\citenamefont {Vizsnyiczai}\ \emph {et~al.}(2017)\citenamefont
  {Vizsnyiczai}, \citenamefont {Frangipane}, \citenamefont {Maggi},
  \citenamefont {Saglimbeni}, \citenamefont {Bianchi},\ and\ \citenamefont
  {Leonardo}}]{vizsnyiczai17}%
  \BibitemOpen
  \bibfield  {author} {\bibinfo {author} {\bibfnamefont {G.}~\bibnamefont
  {Vizsnyiczai}}, \bibinfo {author} {\bibfnamefont {G.}~\bibnamefont
  {Frangipane}}, \bibinfo {author} {\bibfnamefont {C.}~\bibnamefont {Maggi}},
  \bibinfo {author} {\bibfnamefont {F.}~\bibnamefont {Saglimbeni}}, \bibinfo
  {author} {\bibfnamefont {S.}~\bibnamefont {Bianchi}}, \ and\ \bibinfo
  {author} {\bibfnamefont {R.~D.}\ \bibnamefont {Leonardo}},\ }\href@noop {}
  {\bibfield  {journal} {\bibinfo  {journal} {Nat. Comm.}\ }\textbf {\bibinfo
  {volume} {8}},\ \bibinfo {pages} {1} (\bibinfo {year} {2017})}\BibitemShut
  {NoStop}%
\bibitem [{\citenamefont {Reichhardt}\ and\ \citenamefont
  {Reichhardt}(2017)}]{reichhardt17}%
  \BibitemOpen
  \bibfield  {author} {\bibinfo {author} {\bibfnamefont {C.~O.}\ \bibnamefont
  {Reichhardt}}\ and\ \bibinfo {author} {\bibfnamefont {C.}~\bibnamefont
  {Reichhardt}},\ }\href@noop {} {\bibfield  {journal} {\bibinfo  {journal}
  {Annu. Rev. Condens. Matter Phys.}\ }\textbf {\bibinfo {volume} {8}},\
  \bibinfo {pages} {51} (\bibinfo {year} {2017})}\BibitemShut {NoStop}%
\bibitem [{\citenamefont {Pietzonka}\ \emph {et~al.}(2019)\citenamefont
  {Pietzonka}, \citenamefont {Fodor}, \citenamefont {Lohrmann}, \citenamefont
  {Cates},\ and\ \citenamefont {Seifert}}]{pietzonka2019}%
  \BibitemOpen
  \bibfield  {author} {\bibinfo {author} {\bibfnamefont {P.}~\bibnamefont
  {Pietzonka}}, \bibinfo {author} {\bibfnamefont {{\'E}.}~\bibnamefont
  {Fodor}}, \bibinfo {author} {\bibfnamefont {C.}~\bibnamefont {Lohrmann}},
  \bibinfo {author} {\bibfnamefont {M.~E.}\ \bibnamefont {Cates}}, \ and\
  \bibinfo {author} {\bibfnamefont {U.}~\bibnamefont {Seifert}},\ }\href@noop
  {} {\bibfield  {journal} {\bibinfo  {journal} {Phys. Rev. X}\ }\textbf
  {\bibinfo {volume} {9}},\ \bibinfo {pages} {041032} (\bibinfo {year}
  {2019})}\BibitemShut {NoStop}%
\bibitem [{\citenamefont {Krishnamurthy}\ \emph {et~al.}(2016)\citenamefont
  {Krishnamurthy}, \citenamefont {Ghosh}, \citenamefont {Chatterji},
  \citenamefont {Ganapathy},\ and\ \citenamefont {Sood}}]{krishnamurthy2016}%
  \BibitemOpen
  \bibfield  {author} {\bibinfo {author} {\bibfnamefont {S.}~\bibnamefont
  {Krishnamurthy}}, \bibinfo {author} {\bibfnamefont {S.}~\bibnamefont
  {Ghosh}}, \bibinfo {author} {\bibfnamefont {D.}~\bibnamefont {Chatterji}},
  \bibinfo {author} {\bibfnamefont {R.}~\bibnamefont {Ganapathy}}, \ and\
  \bibinfo {author} {\bibfnamefont {A.}~\bibnamefont {Sood}},\ }\href@noop {}
  {\bibfield  {journal} {\bibinfo  {journal} {Nat. Phys.}\ }\textbf {\bibinfo
  {volume} {12}},\ \bibinfo {pages} {1134} (\bibinfo {year}
  {2016})}\BibitemShut {NoStop}%
\bibitem [{\citenamefont {Zakine}\ \emph {et~al.}(2017)\citenamefont {Zakine},
  \citenamefont {Solon}, \citenamefont {Gingrich},\ and\ \citenamefont
  {Van~Wijland}}]{zakine2017}%
  \BibitemOpen
  \bibfield  {author} {\bibinfo {author} {\bibfnamefont {R.}~\bibnamefont
  {Zakine}}, \bibinfo {author} {\bibfnamefont {A.}~\bibnamefont {Solon}},
  \bibinfo {author} {\bibfnamefont {T.}~\bibnamefont {Gingrich}}, \ and\
  \bibinfo {author} {\bibfnamefont {F.}~\bibnamefont {Van~Wijland}},\
  }\href@noop {} {\bibfield  {journal} {\bibinfo  {journal} {Entropy}\ }\textbf
  {\bibinfo {volume} {19}},\ \bibinfo {pages} {193} (\bibinfo {year}
  {2017})}\BibitemShut {NoStop}%
\bibitem [{\citenamefont {Wu}\ and\ \citenamefont {Libchaber}(2000)}]{wu2000}%
  \BibitemOpen
  \bibfield  {author} {\bibinfo {author} {\bibfnamefont {X.-L.}\ \bibnamefont
  {Wu}}\ and\ \bibinfo {author} {\bibfnamefont {A.}~\bibnamefont {Libchaber}},\
  }\href@noop {} {\bibfield  {journal} {\bibinfo  {journal} {Phys. Rev. Lett.}\
  }\textbf {\bibinfo {volume} {84}},\ \bibinfo {pages} {3017} (\bibinfo {year}
  {2000})}\BibitemShut {NoStop}%
\bibitem [{\citenamefont {Martin}\ \emph {et~al.}(2018)\citenamefont {Martin},
  \citenamefont {Nardini}, \citenamefont {Cates},\ and\ \citenamefont
  {Fodor}}]{martin2018}%
  \BibitemOpen
  \bibfield  {author} {\bibinfo {author} {\bibfnamefont {D.}~\bibnamefont
  {Martin}}, \bibinfo {author} {\bibfnamefont {C.}~\bibnamefont {Nardini}},
  \bibinfo {author} {\bibfnamefont {M.~E.}\ \bibnamefont {Cates}}, \ and\
  \bibinfo {author} {\bibfnamefont {{\'E}.}~\bibnamefont {Fodor}},\ }\href@noop
  {} {\bibfield  {journal} {\bibinfo  {journal} {Europhys. Lett.}\ }\textbf
  {\bibinfo {volume} {121}},\ \bibinfo {pages} {60005} (\bibinfo {year}
  {2018})}\BibitemShut {NoStop}%
\bibitem [{\citenamefont {Saha}\ and\ \citenamefont
  {Marathe}(2019)}]{saha2019}%
  \BibitemOpen
  \bibfield  {author} {\bibinfo {author} {\bibfnamefont {A.}~\bibnamefont
  {Saha}}\ and\ \bibinfo {author} {\bibfnamefont {R.}~\bibnamefont {Marathe}},\
  }\href@noop {} {\bibfield  {journal} {\bibinfo  {journal} {J. Stat. Mech.}\
  }\textbf {\bibinfo {volume} {2019}},\ \bibinfo {pages} {094012} (\bibinfo
  {year} {2019})}\BibitemShut {NoStop}%
\bibitem [{\citenamefont {Holubec}\ \emph {et~al.}(2020)\citenamefont
  {Holubec}, \citenamefont {Steffenoni}, \citenamefont {Falasco},\ and\
  \citenamefont {Kroy}}]{holubec2020}%
  \BibitemOpen
  \bibfield  {author} {\bibinfo {author} {\bibfnamefont {V.}~\bibnamefont
  {Holubec}}, \bibinfo {author} {\bibfnamefont {S.}~\bibnamefont {Steffenoni}},
  \bibinfo {author} {\bibfnamefont {G.}~\bibnamefont {Falasco}}, \ and\
  \bibinfo {author} {\bibfnamefont {K.}~\bibnamefont {Kroy}},\ }\href@noop {}
  {\bibfield  {journal} {\bibinfo  {journal} {Phys. Rev. Research}\ }\textbf
  {\bibinfo {volume} {2}},\ \bibinfo {pages} {043262} (\bibinfo {year}
  {2020})}\BibitemShut {NoStop}%
\bibitem [{\citenamefont {Kumari}\ \emph {et~al.}(2020)\citenamefont {Kumari},
  \citenamefont {Pal}, \citenamefont {Saha},\ and\ \citenamefont
  {Lahiri}}]{kumari2020}%
  \BibitemOpen
  \bibfield  {author} {\bibinfo {author} {\bibfnamefont {A.}~\bibnamefont
  {Kumari}}, \bibinfo {author} {\bibfnamefont {P.}~\bibnamefont {Pal}},
  \bibinfo {author} {\bibfnamefont {A.}~\bibnamefont {Saha}}, \ and\ \bibinfo
  {author} {\bibfnamefont {S.}~\bibnamefont {Lahiri}},\ }\href@noop {}
  {\bibfield  {journal} {\bibinfo  {journal} {Phys. Rev. E}\ }\textbf {\bibinfo
  {volume} {101}},\ \bibinfo {pages} {032109} (\bibinfo {year}
  {2020})}\BibitemShut {NoStop}%
\bibitem [{\citenamefont {Ekeh}\ \emph {et~al.}(2020)\citenamefont {Ekeh},
  \citenamefont {Cates},\ and\ \citenamefont {Fodor}}]{ekeh2020}%
  \BibitemOpen
  \bibfield  {author} {\bibinfo {author} {\bibfnamefont {T.}~\bibnamefont
  {Ekeh}}, \bibinfo {author} {\bibfnamefont {M.~E.}\ \bibnamefont {Cates}}, \
  and\ \bibinfo {author} {\bibfnamefont {E.}~\bibnamefont {Fodor}},\
  }\href@noop {} {\bibfield  {journal} {\bibinfo  {journal} {Phys. Rev. E}\
  }\textbf {\bibinfo {volume} {102}},\ \bibinfo {pages} {010101(R)} (\bibinfo
  {year} {2020})}\BibitemShut {NoStop}%
\bibitem [{\citenamefont {Marconi}\ \emph {et~al.}(2017)\citenamefont
  {Marconi}, \citenamefont {Puglisi},\ and\ \citenamefont {Maggi}}]{umb17}%
  \BibitemOpen
  \bibfield  {author} {\bibinfo {author} {\bibfnamefont {U.~M.~B.}\
  \bibnamefont {Marconi}}, \bibinfo {author} {\bibfnamefont {A.}~\bibnamefont
  {Puglisi}}, \ and\ \bibinfo {author} {\bibfnamefont {C.}~\bibnamefont
  {Maggi}},\ }\href@noop {} {\bibfield  {journal} {\bibinfo  {journal}
  {Scientific Reports}\ }\textbf {\bibinfo {volume} {7}},\ \bibinfo {pages}
  {46496} (\bibinfo {year} {2017})}\BibitemShut {NoStop}%
\bibitem [{\citenamefont {Puglisi}\ and\ \citenamefont
  {Marconi}(2017)}]{clau17}%
  \BibitemOpen
  \bibfield  {author} {\bibinfo {author} {\bibfnamefont {A.}~\bibnamefont
  {Puglisi}}\ and\ \bibinfo {author} {\bibfnamefont {U.~M.~B.}\ \bibnamefont
  {Marconi}},\ }\href@noop {} {\bibfield  {journal} {\bibinfo  {journal}
  {Entropy}\ }\textbf {\bibinfo {volume} {19}},\ \bibinfo {pages} {356}
  (\bibinfo {year} {2017})}\BibitemShut {NoStop}%
\bibitem [{\citenamefont {Arlt}\ \emph {et~al.}(2019)\citenamefont {Arlt},
  \citenamefont {Martinez}, \citenamefont {Dawson}, \citenamefont {Pilizota},\
  and\ \citenamefont {Poon}}]{arlt2019dynamics}%
  \BibitemOpen
  \bibfield  {author} {\bibinfo {author} {\bibfnamefont {J.}~\bibnamefont
  {Arlt}}, \bibinfo {author} {\bibfnamefont {V.~A.}\ \bibnamefont {Martinez}},
  \bibinfo {author} {\bibfnamefont {A.}~\bibnamefont {Dawson}}, \bibinfo
  {author} {\bibfnamefont {T.}~\bibnamefont {Pilizota}}, \ and\ \bibinfo
  {author} {\bibfnamefont {W.~C.}\ \bibnamefont {Poon}},\ }\href@noop {}
  {\bibfield  {journal} {\bibinfo  {journal} {Nat. Comm.}\ }\textbf {\bibinfo
  {volume} {10}},\ \bibinfo {pages} {1} (\bibinfo {year} {2019})}\BibitemShut
  {NoStop}%
\bibitem [{\citenamefont {Schmidt}\ \emph {et~al.}(2019)\citenamefont
  {Schmidt}, \citenamefont {Liebchen}, \citenamefont {L{\"o}wen},\ and\
  \citenamefont {Volpe}}]{schmidt2019light}%
  \BibitemOpen
  \bibfield  {author} {\bibinfo {author} {\bibfnamefont {F.}~\bibnamefont
  {Schmidt}}, \bibinfo {author} {\bibfnamefont {B.}~\bibnamefont {Liebchen}},
  \bibinfo {author} {\bibfnamefont {H.}~\bibnamefont {L{\"o}wen}}, \ and\
  \bibinfo {author} {\bibfnamefont {G.}~\bibnamefont {Volpe}},\ }\href@noop {}
  {\bibfield  {journal} {\bibinfo  {journal} {J. Chem. Phys.}\ }\textbf
  {\bibinfo {volume} {150}},\ \bibinfo {pages} {094905} (\bibinfo {year}
  {2019})}\BibitemShut {NoStop}%
\bibitem [{\citenamefont {Brandner}\ \emph {et~al.}(2015)\citenamefont
  {Brandner}, \citenamefont {Saito},\ and\ \citenamefont
  {Seifert}}]{brandner2015}%
  \BibitemOpen
  \bibfield  {author} {\bibinfo {author} {\bibfnamefont {K.}~\bibnamefont
  {Brandner}}, \bibinfo {author} {\bibfnamefont {K.}~\bibnamefont {Saito}}, \
  and\ \bibinfo {author} {\bibfnamefont {U.}~\bibnamefont {Seifert}},\
  }\href@noop {} {\bibfield  {journal} {\bibinfo  {journal} {Phys. Rev. X}\
  }\textbf {\bibinfo {volume} {5}},\ \bibinfo {pages} {031019} (\bibinfo {year}
  {2015})}\BibitemShut {NoStop}%
\bibitem [{\citenamefont {Seifert}(2012)}]{seifertrev}%
  \BibitemOpen
  \bibfield  {author} {\bibinfo {author} {\bibfnamefont {U.}~\bibnamefont
  {Seifert}},\ }\href@noop {} {\bibfield  {journal} {\bibinfo  {journal} {Rep.
  Prog. Phys.}\ }\textbf {\bibinfo {volume} {75}},\ \bibinfo {pages} {126001}
  (\bibinfo {year} {2012})}\BibitemShut {NoStop}%
\bibitem [{\citenamefont {Jarzynski}(2007)}]{jarzynski2007comparison}%
  \BibitemOpen
  \bibfield  {author} {\bibinfo {author} {\bibfnamefont {C.}~\bibnamefont
  {Jarzynski}},\ }\href@noop {} {\bibfield  {journal} {\bibinfo  {journal} {C R
  Phys.}\ }\textbf {\bibinfo {volume} {8}},\ \bibinfo {pages} {495} (\bibinfo
  {year} {2007})}\BibitemShut {NoStop}%
\bibitem [{Note2()}]{Note2}%
  \BibitemOpen
  \bibinfo {note} {A recent different definition for $Q_h$ has been proposed in
  the context of active engines coupled both with a steady active bath and a
  steady thermal bath which are of course at different temperatures~\cite
  {fodor2021active}: in that case a cyclical engine can be obtained by tuning
  in time {\protect \em two} parameters of the external potential and the
  proposed definition of adsorbed heat is the whole heat exchanged with the
  active bath, which is positive on average. In our opinion this choice should
  be debated, as it implies that no heat is dissipated in the active bath
  itself, a fact which we are challenging.}\BibitemShut {Stop}%
\bibitem [{Note3()}]{Note3}%
  \BibitemOpen
  \bibinfo {note} {With this definition the entropy produced in a period due to
  work and heat flux are equal to $S_{prod,W}=\beta _c W_p$ and
  $S_{prod,h}=Q_h(\beta _c-\beta _h)$ respectively~\cite {brandner2015}, so
  that $\eta =-W_p/Q_h = S_{prod,W}(\beta _h-\beta _c)/(S_{prod,h}\beta _c)$.
  In the quasi-static limit $S_{prod,W}/S_{prod,h}=-1$, which leads to the
  Carnot efficiency $\eta =1-\beta _h/\beta _c=\eta _c$.}\BibitemShut {Stop}%
\bibitem [{\citenamefont {Maggi}\ \emph {et~al.}(2015)\citenamefont {Maggi},
  \citenamefont {Marconi}, \citenamefont {Gnan},\ and\ \citenamefont {{{Di
  Leonardo}}}}]{MBGL15}%
  \BibitemOpen
  \bibfield  {author} {\bibinfo {author} {\bibfnamefont {C.}~\bibnamefont
  {Maggi}}, \bibinfo {author} {\bibfnamefont {U.~M.~B.}\ \bibnamefont
  {Marconi}}, \bibinfo {author} {\bibfnamefont {N.}~\bibnamefont {Gnan}}, \
  and\ \bibinfo {author} {\bibfnamefont {R.}~\bibnamefont {{{Di Leonardo}}}},\
  }\href@noop {} {\bibfield  {journal} {\bibinfo  {journal} {Scientific
  Reports}\ }\textbf {\bibinfo {volume} {5}},\ \bibinfo {pages} {10742}
  (\bibinfo {year} {2015})}\BibitemShut {NoStop}%
\bibitem [{\citenamefont {Caprini}\ and\ \citenamefont {Marini
  Bettolo~Marconi}(2021)}]{caprini21}%
  \BibitemOpen
  \bibfield  {author} {\bibinfo {author} {\bibfnamefont {L.}~\bibnamefont
  {Caprini}}\ and\ \bibinfo {author} {\bibfnamefont {U.}~\bibnamefont {Marini
  Bettolo~Marconi}},\ }\href@noop {} {\bibfield  {journal} {\bibinfo  {journal}
  {J.Chem. Phys.}\ }\textbf {\bibinfo {volume} {154}},\ \bibinfo {pages}
  {024902} (\bibinfo {year} {2021})}\BibitemShut {NoStop}%
\bibitem [{\citenamefont {Takatori}\ \emph {et~al.}(2014)\citenamefont
  {Takatori}, \citenamefont {Yan},\ and\ \citenamefont
  {Brady}}]{takatori2014swim}%
  \BibitemOpen
  \bibfield  {author} {\bibinfo {author} {\bibfnamefont {S.~C.}\ \bibnamefont
  {Takatori}}, \bibinfo {author} {\bibfnamefont {W.}~\bibnamefont {Yan}}, \
  and\ \bibinfo {author} {\bibfnamefont {J.~F.}\ \bibnamefont {Brady}},\
  }\href@noop {} {\bibfield  {journal} {\bibinfo  {journal} {Phys. Rev. Lett.}\
  }\textbf {\bibinfo {volume} {113}},\ \bibinfo {pages} {028103} (\bibinfo
  {year} {2014})}\BibitemShut {NoStop}%
\bibitem [{\citenamefont {Winkler}\ \emph {et~al.}(2015)\citenamefont
  {Winkler}, \citenamefont {Wysocki},\ and\ \citenamefont
  {Gompper}}]{winkler2015virial}%
  \BibitemOpen
  \bibfield  {author} {\bibinfo {author} {\bibfnamefont {R.~G.}\ \bibnamefont
  {Winkler}}, \bibinfo {author} {\bibfnamefont {A.}~\bibnamefont {Wysocki}}, \
  and\ \bibinfo {author} {\bibfnamefont {G.}~\bibnamefont {Gompper}},\
  }\href@noop {} {\bibfield  {journal} {\bibinfo  {journal} {Soft Matter}\
  }\textbf {\bibinfo {volume} {11}},\ \bibinfo {pages} {6680} (\bibinfo {year}
  {2015})}\BibitemShut {NoStop}%
\bibitem [{\citenamefont {Jung}\ and\ \citenamefont
  {H\"anggi}(1987)}]{hanggi87}%
  \BibitemOpen
  \bibfield  {author} {\bibinfo {author} {\bibfnamefont {P.}~\bibnamefont
  {Jung}}\ and\ \bibinfo {author} {\bibfnamefont {P.}~\bibnamefont
  {H\"anggi}},\ }\href@noop {} {\bibfield  {journal} {\bibinfo  {journal}
  {Phys. Rev. A}\ }\textbf {\bibinfo {volume} {35}},\ \bibinfo {pages} {4464}
  (\bibinfo {year} {1987})}\BibitemShut {NoStop}%
\bibitem [{\citenamefont {H\"anggi}\ and\ \citenamefont
  {Jung}(1995)}]{hanggi95}%
  \BibitemOpen
  \bibfield  {author} {\bibinfo {author} {\bibfnamefont {P.}~\bibnamefont
  {H\"anggi}}\ and\ \bibinfo {author} {\bibfnamefont {P.}~\bibnamefont
  {Jung}},\ }\href@noop {} {\bibfield  {journal} {\bibinfo  {journal} {Adv.
  Chem. Phys.}\ }\textbf {\bibinfo {volume} {89}},\ \bibinfo {pages} {239}
  (\bibinfo {year} {1995})}\BibitemShut {NoStop}%
\bibitem [{\citenamefont {Izumida}\ and\ \citenamefont
  {Okuda}(2010)}]{izumida}%
  \BibitemOpen
  \bibfield  {author} {\bibinfo {author} {\bibfnamefont {Y.}~\bibnamefont
  {Izumida}}\ and\ \bibinfo {author} {\bibfnamefont {K.}~\bibnamefont
  {Okuda}},\ }\href@noop {} {\bibfield  {journal} {\bibinfo  {journal} {Eur.
  Phys. J. B}\ }\textbf {\bibinfo {volume} {77}},\ \bibinfo {pages} {499}
  (\bibinfo {year} {2010})}\BibitemShut {NoStop}%
\bibitem [{\citenamefont {Callen}(1985)}]{callen}%
  \BibitemOpen
  \bibfield  {author} {\bibinfo {author} {\bibfnamefont {H.~B.}\ \bibnamefont
  {Callen}},\ }\href@noop {} {\emph {\bibinfo {title} {Thermodynamics and an
  Introduction to Thermostatics}}}\ (\bibinfo  {publisher} {Wiley},\ \bibinfo
  {year} {1985})\BibitemShut {NoStop}%
\bibitem [{\citenamefont {Curzon}\ and\ \citenamefont
  {Ahlborn}(1975)}]{curzon1975}%
  \BibitemOpen
  \bibfield  {author} {\bibinfo {author} {\bibfnamefont {F.~L.}\ \bibnamefont
  {Curzon}}\ and\ \bibinfo {author} {\bibfnamefont {B.}~\bibnamefont
  {Ahlborn}},\ }\href@noop {} {\bibfield  {journal} {\bibinfo  {journal} {Am.
  J. Phys.}\ }\textbf {\bibinfo {volume} {43}},\ \bibinfo {pages} {22}
  (\bibinfo {year} {1975})}\BibitemShut {NoStop}%
\bibitem [{Note4()}]{Note4}%
  \BibitemOpen
  \bibinfo {note} {Our numerical scheme is a classical 4-th order Runge-Kutta
  integrator with time-step $dt=10^{-3}$ for the passive system and
  $dt=10^{-4}$ for the active one.}\BibitemShut {Stop}%
\bibitem [{\citenamefont {Fodor}\ and\ \citenamefont
  {Cates}(2021)}]{fodor2021active}%
  \BibitemOpen
  \bibfield  {author} {\bibinfo {author} {\bibfnamefont {{\'E}.}~\bibnamefont
  {Fodor}}\ and\ \bibinfo {author} {\bibfnamefont {M.~E.}\ \bibnamefont
  {Cates}},\ }\href@noop {} {\enquote {\bibinfo {title} {Active engines:
  Thermodynamics moves forward},}\ } (\bibinfo {year} {2021})\BibitemShut
  {NoStop}%
\end{thebibliography}%

\end{document}